\documentclass[twocolumn,showpacs,superscriptaddress,floatfix,prb]{revtex4-2}
\usepackage{graphicx,amsfonts,amssymb,amsmath,hyperref,hypcap,enumerate}

\usepackage{amsthm}
\usepackage{multirow}
\usepackage{graphicx}
\usepackage{dcolumn}   % needed for some tables
\usepackage{bm}        % for math
\usepackage{amssymb}   % for math
\usepackage{amsmath}
\usepackage{epstopdf}
\usepackage{color}
\usepackage{subfigure}
\usepackage{diagbox}

\usepackage{multirow}
\usepackage{soul}
\begin{document}
	\title{Universality classes of the Anderson transitions 
		driven by quasiperiodic potential\\
	in the three-dimensional Wigner-Dyson symmetry classes}
	
	%Universality classes of the Anderson transitions 
	%in the three-dimensional Anderson, Peierls-phase, and Ando models
	%driven by quasiperiodic potential
	
	\author{Xunlong Luo}
	\email{luoxunlong@pku.edu.cn}
	\affiliation{Science and Technology on Surface Physics and Chemistry Laboratory, Mianyang 621907, China}
	
	\author{Tomi Ohtsuki}
	\email{ohtsuki@sophia.ac.jp}
	\affiliation{Physics Division, Sophia University, Chiyoda-ku, Tokyo 102-8554, Japan}
	
	%\author{Ryuichi Shindou}
	%\email{rshindou@pku.edu.cn}
	%\affiliation{International Center for Quantum Materials, Peking University, Beijing 100871, China}
	%\affiliation{Collaborative Innovation Center of Quantum Matter, Beijing 100871, China}
	
	\date{\today}
	\begin{abstract}
		Quasiperiodic system is an intermediate state %\textcolor{blue}{material$\rightarrow$?} 
		between periodic and disordered systems
		with unique delocalization-localization transition driven by the
		quasiperiodic potential (QP).  
		One of the intriguing questions is whether the universality class of the
		Anderson transition (AT) driven by QP is similar to 
		that of the AT driven by the random potential in the same symmetry class. 
		Here,  we study the critical behavior of the ATs driven by QP in the
		three-dimensional (3D) Anderson model, Peierls phase model, and Ando model,
		which belong to the Wigner-Dyson symmetry classes.
		The localization length and two-terminal conductance have been calculated by the transfer matrix method,
		and we argue that their error estimations in statistics suffer from the correlation of QP.
		With the correlation under control, the critical exponents $\nu$ of the ATs driven by QP
		are estimated by the finite size scaling analysis of conductance,
		which are consistent with $\nu$'s of the ATs driven by the random potential. 
		Moreover, the critical conductance distribution and the level spacing ratio distribution have been studied. 
		We also find that a convolutional neural network trained by the localized/delocalized
		wavefunctions in a disordered system predicts the localized/delocalized wavefunctions in quasiperiodic systems.
		Our numerical results strongly support that the universality classes of the ATs driven
		by QP and random potential are similar in the 3D Wigner-Dyson symmetry classes.
	\end{abstract}
	\maketitle

	\section{Introduction}
	In the seminal work by Anderson, the extended wavefunction could become localized 
	driven by random potential because of the quantum interference \cite{Anderson58}. 
	This delocalization-localization transition is called the Anderson transition (AT),
	which can be described by the single parameter scaling theory \cite{Abrahams79}.
	As a second-order phase transition, the universality class of AT, which is characterized by the 
	critical exponents, depends on the symmetry and dimension, 
	but not the detail of the system \cite{Evers08}.
	
	On the other hand, the quasiperiodic potential (QP) can also drive the Anderson localization,
	which has been extensively studied both theoretically
	\cite{Aubry80,Sokoloff80,Suslov82,Kohmoto83,  %1D
		Soukoulis82,Sarma88,Biddle10,Yao19,Ganeshan15,Roy21, % mobility edge
		Sanchez00,Guo14,Rossignolo19,Devakul17,Sutradhar19,Huang19,Szabo20,% 2D/3D 
		Goncalves21,Huang16,  % 2D and 3D added
		Wang17PRA,Pixley18}   % topology
	and experimentally \cite{Lahini09,Sbroscia20,Wang19,  %   optical lattice
		Roati08,Lopez12,Luschen18,               %   cold atome
		Goblot20,An21}.                               %   mobility edge
	Quasiperiodic systems are ubiquitous in nature and can be realized in 
	the quasicrystals \cite{Shechtman84}, photonic lattices \cite{Lahini09,Sbroscia20,Wang19}, 
	ultracold quantum gases \cite{Roati08,Lopez12,Luschen18,Schreiber15,Bordia17,Deissler10,Nakajima21}, and Moir\'{e} superlattices
	(i.e., the twisted bilayer graphene at $30^{\circ}$\cite{Bistritzer11,Yao18PNAS}).
	One of the famous examples is the one-dimensional (1D) Aubry-Andr\'{e}-Harper (AAH) model \cite{Harper55,Aubry80}, 
	where there is a delocalization-localization transition with the same critical point
	at any Fermi energy due to the self-duality.
	Generalized AAH models involve the exact mobility edges \cite{Soukoulis82,Sarma88,Biddle10,Yao19,Ganeshan15,Roy21}, 
    connections to higher-dimensional topological matter phases \cite{Hofstadter76,Kraus12,Kraus13,Verbin13,Zilberberg18}, 
	and higher-dimensional ATs \cite{Devakul17,Huang19,Sutradhar19,Szabo20}.
	
	\begin{table*}[tb]
		\centering
		%	\footnotesize
		\setlength{\tabcolsep}{3mm}
		\caption{Polynomial fitting results for the two-terminal conductance around 
			the Anderson transition points for three dimensional Anderson model (AM), Peierls phase model (PPM)
			and Ando model with quasiperiodic potential. The goodness of fit (GOF), critical quasiperiodic potential
			strength $V_{c}$, critical exponent $\nu$, and the scaling dimension of the least
			irrelevant scaling variable $-y$ are 	shown for various system sizes and for different 
			orders of the Taylor expansion: $(m_1, n_1, m_2, n_2)$. 
			The square bracket is the 95\% confidence interval.
		}
		\begin{tabular}{cccccccccc}
			\hline
			\hline	
			Model&	$L$&$m_1$& $n_1$ & $m_2$ & $n_2$ & GOF & $V_c$ & $\nu$ & $y$ \\
			\hline
			AM&	10-24&3&3&0&1&0.106&2.229[2.226, 2.233]&1.599[1.569, 1.637]&3.9[3.3, 4.6]\\
			&	10-28&3&3&0&1&0.101&2.233[2.230, 2.236]&1.613[1.587, 1.639]&3.5[3.0, 4.1]\\
			\hline
			PPM&10-24&3&3&0&1&0.105&2.335[2.331, 2.341]&1.43[1.39, 1.49]&3.0[2.2, 3.8]\\ 
			&	10-28&3&3&0&1&0.162&2.339[2.336, 2.342]&1.45[1.42, 1.49]&2.8[2.3, 3.4]\\ 
			\hline
			Ando&	10-20&2&3&0&1&0.112&2.449[2.445, 2.455]&1.37[1.35, 1.39]&4.3[3.6, 4.9]\\
			&10-20&3&3&0&1&0.139&2.449[2.445, 2.455]&1.36[1.34, 1.38]&4.3[3.7, 4.9]\\
			\hline
			\hline	
		\end{tabular}
		\label{table_nu}
	\end{table*}

	Both random potential and QP can drive the ATs, 
	and a natural question is whether the universality classes of the ATs driven 
	by QP and those by random potential are the same or not \cite{Devakul17}.
	For 1D and two-dimensional (2D) systems in orthogonal class, the infinitesimal random potential 
	will localize the wavefunctions and there is no AT at all \cite{Abrahams79}.
	However, the 1D AAH model shows a delocalization-localization transition driven by QP with exact critical exponent $\nu=1$
	\cite{Aubry80}, and the critical exponent is estimated as $\nu\approx1$ \cite{Huang19}
	for the generalized 2D AAH model.
	Both of them belong to the orthogonal class.
	Therefore the ATs in 1D and 2D AAH models belong to new universality classes without counterparts in the
	disordered systems. However, it is reported that the universality class of AT in the generalized three-dimensional (3D)  AAH model 
	is similar to that of the AT driven by random potential in 3D orthogonal class \cite{Devakul17,Sutradhar19}. 
	Considering the uniqueness of 3D case,
	it is necessary to study the critical behavior by independent methods, 
	such as the localization length, conductance, level statistics, and deep learning.
	Moreover, it is also interesting to check the conclusion in other symmetry classes, 
	such as unitary and symplectic symmetry classes \cite{Wigner51,Dyson62,Dyson62TFW}.

	%The localization length on quasi-one-dimensional bars has been calculated 
	%to extract the critical exponent $\nu$ just after the proposal of the scaling theory for Anderson localization \cite{Pichard81,MacKinnon81,MacKinnon83}.
	%On the other hand, the conductance is the relevant parameter in the single parameter scaling theory for the AT
	%proposed in the pioneering work \cite{Abrahams79}, which has also been used to extract the critical behavior successfully \cite{Slevin01,Slevin03}.
	
	In this paper, we focus on the critical behavior of the ATs driven by QP in the 3D Anderson model (AM) \cite{Anderson58},
	Peierls phase model (PPM)~\cite{Peierls33,Luttinger51},
	and Ando model \cite{Ando89,Kawarabayashi96}, which belong to orthogonal, unitary, and symplectic classes, respectively. 
	The localization length and conductance have been calculated by the transfer matrix method \cite{Pichard81,MacKinnon83,MacKinnon94}.
	We find that the estimates of their statistical error are difficult
	since the samples are not independent due to QP.
	Because the estimates of critical exponents $\nu$ by the polynomial fitting method are 
	sensitive to the error bar of the physical quantities \cite{Slevin14},
	the incorrect error bar for the localization length or conductance will result in incorrect estimates of $\nu$.
	We have chosen the proper number of samples so that they are regarded as independent,
	and the critical exponents $\nu$ have been estimated based on the conductance,
	which are similar to those of the ATs driven by random potential (see Table \ref{table_nu}).
	Based on the critical points estimated by conductance, we also calculated the critical conductance distributions for the three models,
	which are independent of the system size and significantly different from each other.
	By properly taking account of the anisotropy, they are shown to be the same as those of disordered systems in the same symmetry class.
	Although the level spacing ratio distribution $P(r)$ in the delocalized phase of 1D AAH model is special \cite{Li16,Deng17}, 
	we find that $P(r)$ in the delocalized states of the three models in this paper follow the Wigner-like surmises  \cite{Atas13}.
	To further understand the localization properties, we constructed a convolutional neural network (CNN), 
	whose parameters have been trained by the localized/delocalized wavefunctions in a disordered system \cite{Ohtsuki20}. 
	We find this CNN can predict the localized/delocalized wavefunctions in quasiperiodic systems,
	which indicates that the localization properties in the disordered and quasiperiodic systems are similar.
	
	The rest of the paper is organized as follows.
	In the next section, we will first introduce the three models with QP belonging to the Wigner-Dyson symmetry classes. 
	Next, we will study the phase diagram spanned by eigenenergy and QP strength, and the mobility edge in section \ref{sec_PD_ME}. 
	Then we will discuss the finite size scaling of conductance in section \ref{sec_FSS}.
	Moreover, the critical conductance distribution (section \ref{sec_CCD}) 
	and the level spacing ratio distribution (section \ref{sec_Pr}) are studied.
	In section \ref{sec_CNN}, we apply the CNN trained for the random potential to the wavefunctions in QP, 
	and show that the CNN correctly predicts the critical strength of QP.
	The final section is devoted to the summary and concluding remarks.
	
	\begin{figure*}[bt]
		\centering
		\subfigure[ Anderson model]{
			\begin{minipage}[t]{0.333\linewidth}
				\centering
				\includegraphics[width=1\linewidth]{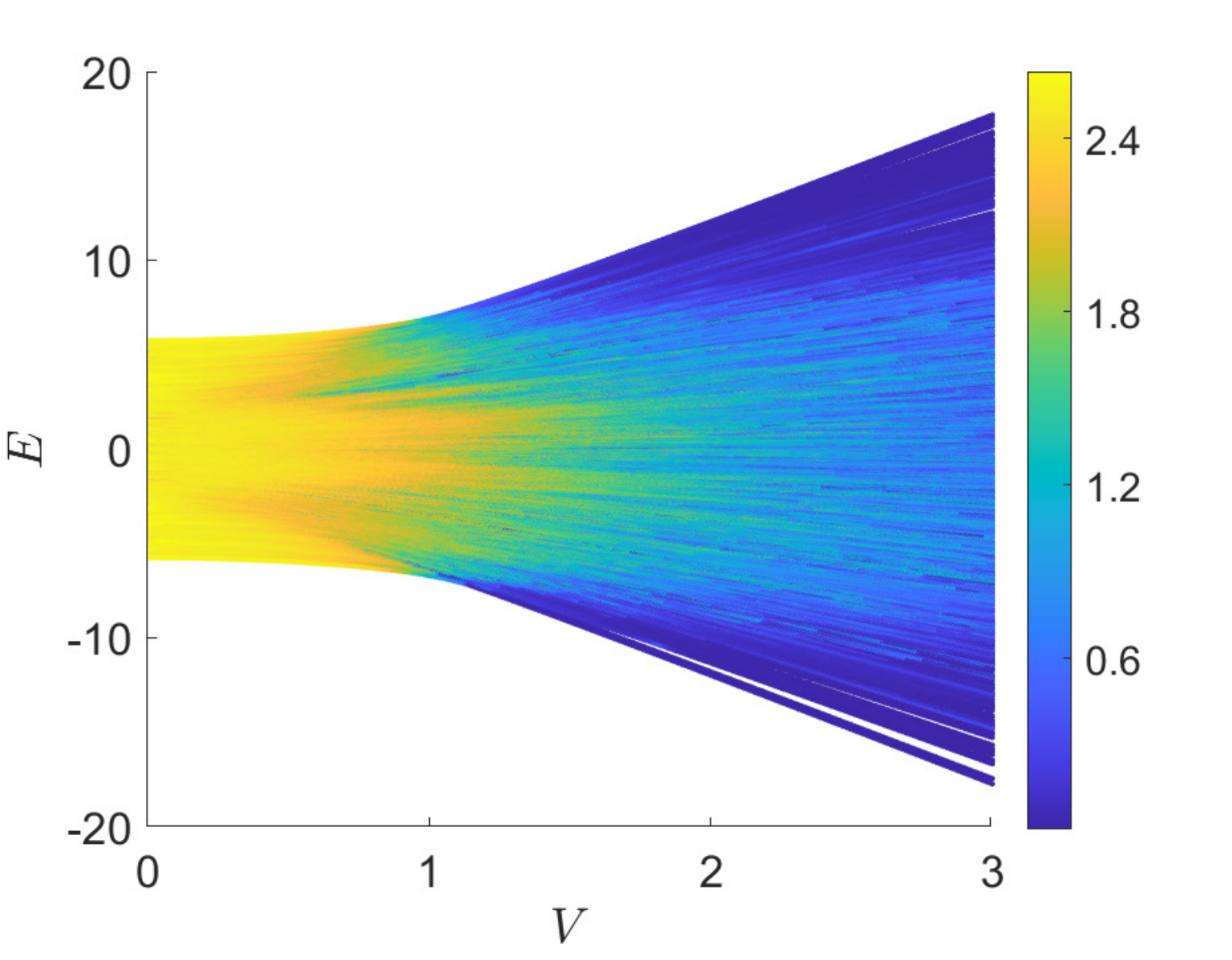}
				%{3D_AAH_couple_model_FBC_L10}
				%\caption{fig1}
			\end{minipage}%
		}%
		\subfigure[Peierls phase model ]{
			\begin{minipage}[t]{0.333\linewidth}
				\centering
				\includegraphics[width=1\linewidth]{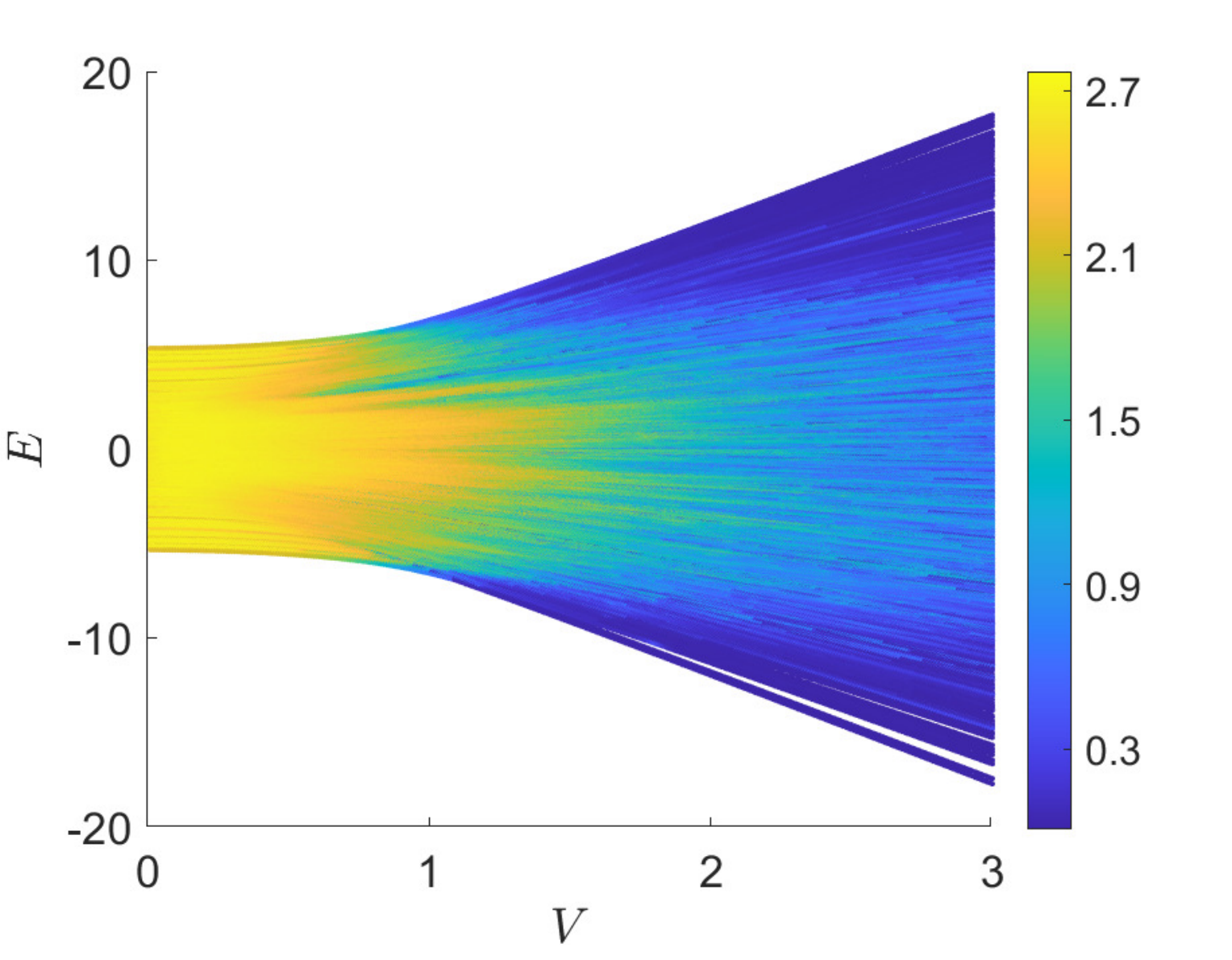}
				%{PPM_3D_couple_L10 }
				%\caption{fig1}
			\end{minipage}%
		}%
		\subfigure[Ando model]{
			\begin{minipage}[t]{0.333\linewidth}
				\centering
				\includegraphics[width=1\linewidth]{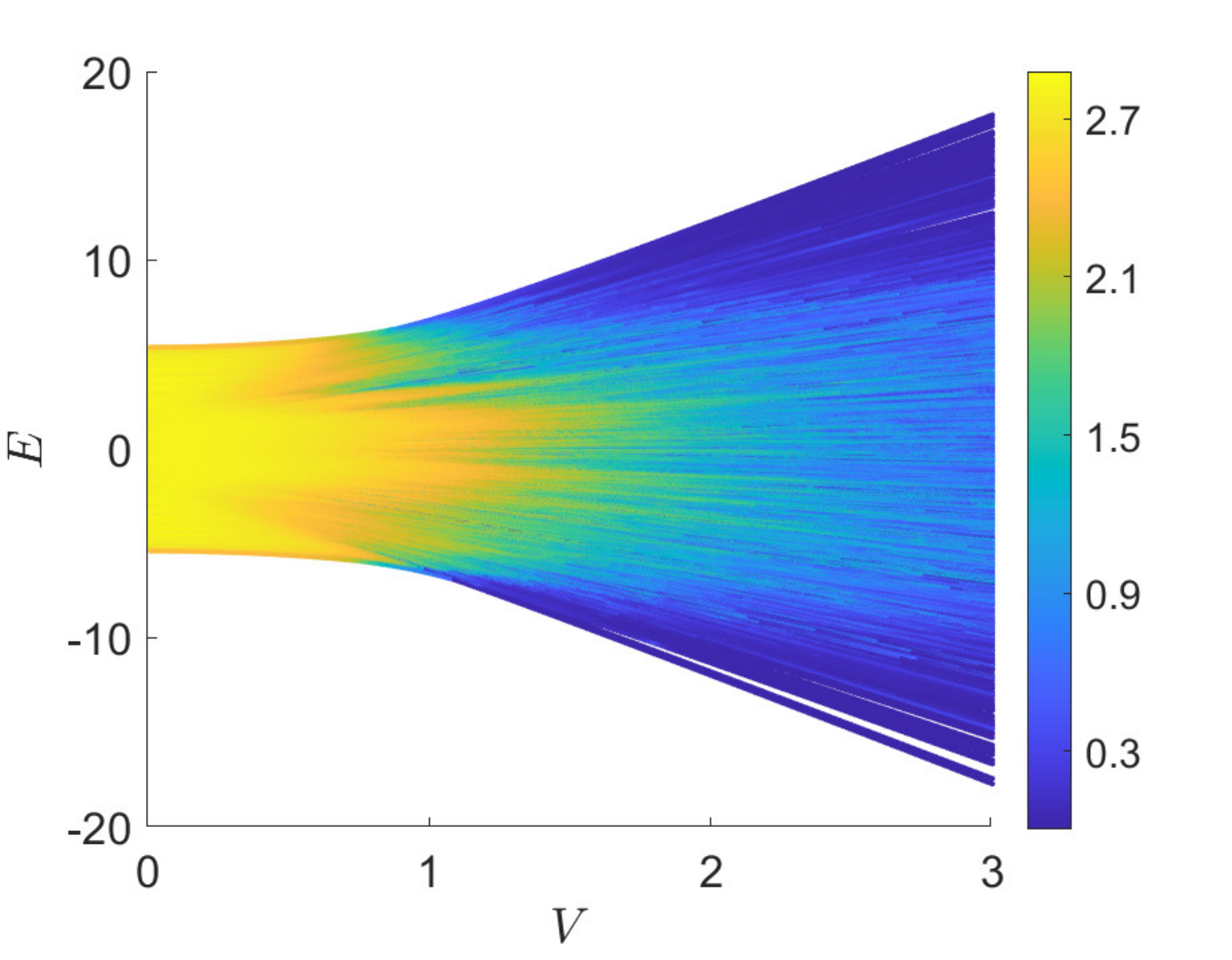}
				%{ Ando_3D_couple_L10 }
				%\caption{fig1}
			\end{minipage}%
		}%	
		\caption{Normalized participation ratio in the logarithmic form $\eta$
			for different eigenenergies $E$ as a function of quasiperiodic potential strength $V$ for the three quasiperiodic models.
			$\eta\approx3$ stands for the well delocalized state and $\eta\approx 0$ for the strongly localized state. 
			$\eta$ is simulated with cubic system with linear size $L=10$.}
		\label{fig_D2}
	\end{figure*}
	
	\section{Model and quasiperiodic potential}\label{sec_model}
	We first introduce the following tight-binding models defined on a 3D
	cubic lattice,
	\begin{align}
		{\cal H} = \sum_{\bm r} \varepsilon_{\bm r} c^{\dagger}_{\bm r} c_{\bm r} + \sum_{\langle {\bm r},{\bm r'}\rangle} 
		e^{2\pi {\rm i} \theta_{{\bm r},{\bm r'}}} c^{\dagger}_{\bm r} c_{\bm r'},  
		\label{O1U1} 
	\end{align}
	where $c^{\dagger}_{\bm r}$ ($c_{\bm r}$) is the creation (annihilation) operator. 
	$\bm r$ and $\bm r'$ specify the cubic lattice site; ${\bm r}=(r_x,r_y,r_z)$ and $r_{\mu}=1,2,\cdots,L_{\mu}$ with $\mu=x,y,z$.
	$\langle {\bm r},{\bm r'}\rangle$ means that $\bm r$ and $\bm r'$ are the nearest neighbor lattice sites. 
	$\rm i=\sqrt{-1}$ is the  imaginary unit.
	The model in Eq. (\ref{O1U1}) is the Anderson model with $\theta_{{\bm r},{\bm r'}}$ = 0 \cite{Anderson58}.
	When $\theta_{{\bm r},{\bm r}+{\bm e}_z} = \Phi \!\ r_x$ with ${\bm e}_z$ the unit vector in $z$ direction
	and $\theta_{{\bm r},{\bm r'}}$ = 0 in the hoppings along $x$ and $y$ directions, 
	the model is the Peierls phase model~\cite{Peierls33,Luttinger51}.
	$\Phi$ is a magnetic gauge flux that penetrates through every square plaquette in the $z$-$x$ plane of the cubic lattice in units of $h/e$, 
	where $h$ is the Planck constant and $e$ is the elementary charge.
	We set $\Phi=0.1$ in the following calculations.
	
	The Ando model can be constructed in the 3D cubic lattice as follows \cite{Ando89,Kawarabayashi96},
	\begin{align}
		{\cal H}=\sum_{\bm r} \varepsilon_{\bm r} c_{{\bm r},\sigma}^{\dagger}c_{{\bm r},\sigma}
		+\sum_{\langle {\bm r}, {\bm r'}\rangle,\sigma,\sigma'}R({\bm r},{\bm r'})_{\sigma,\sigma'}
		c_{{\bm r},\sigma}^{\dagger}c_{{\bm r'},\sigma'}, \label{symplectic}
	\end{align}
	where $c_{{\bm r},\sigma}^{\dagger}$($c_{{\bm r},\sigma}$) is the creation (annihilation) operator with $\sigma=\uparrow,\downarrow$.
	We set the spin-orbit coupling between the nearest neighbor lattice sites as
	$R({\bm r}, {\bm r}+{\bm e}_{\mu})=e^{{\rm i}\theta\sigma_{\mu}}$,
	where ${\bm e}_{\mu}$ is the unit vector in the $\mu=x,y,z$ directions and $\sigma_{\mu}$ is
	the Pauli spin matrix. 
	The parameter $\theta$, which is set to be $\theta = \pi/6$ in our simulations, represents the strength of spin-orbit coupling.

	$\varepsilon_{\bm r}$ in Eq. (\ref{O1U1}) and (\ref{symplectic}) is the onsite potential.
	Here we study the critical behavior of ATs driven by QP, so
	the QP is added into these three models instead of random potential.
	QP has already been directly realized in %quasicrystals \cite{Shechtman84}, 
	the photonic lattices \cite{Lahini09,Sbroscia20},
	ultracold atomic experiments \cite{Roati08,Deissler10,Lopez12,Schreiber15,Bordia17,Luschen18,Nakajima21}, 
	and Moir\'{e} superlattices \cite{Bistritzer11,Yao18PNAS}.
	Following Ref. \onlinecite{Devakul17}, we set
	\begin{align}
		\varepsilon_{\bm r}=2V\sum_{i=1}^{3} \cos(2\pi {\bm b}_i\cdot {\bm r} +\phi_i), \label{epsilon}
	\end{align}
	where $V$ is the strength of  QP, $\{ {\bm b}_i \}$ are vectors determining the quasiperiodicity, 
	and $\{\phi_i\}$  are the arbitrary phases. 
	We write ${\bm b}_i$ in the matrix form 
	${\bm B}=[{\bm b}_1\ \ {\bm b}_2\ \ {\bm b}_3]^{\rm T}$.
	In order to keep quasiperiodicity and make the model nonseparable, 
	we set $\bm B=\beta  {\bm R}(\alpha)$ with $\beta=\frac{\sqrt{5}-1}{2}$ and 
	\begin{align}\label{matrixB}
		{\bm R}(\alpha)=
		\begin{bmatrix}
			c^2+s^3& cs  &cs^2-cs\\
			cs    &-s   &c^2\\
			cs^2-cs& c^2&  c^2s+s^2 \\
		\end{bmatrix},
	\end{align}
	where $c=\cos(\alpha)$ and $s=\sin(\alpha)$. 
	$\beta=\frac{\sqrt{5}-1}{2}$ is the golden mean, which is an irrational number characterizing the quasiperiodicity.
	$\bm R(\alpha)$ is a symmetric orthonormal matrix which makes the model nonseparable,
	as long as we avoid $\alpha = 0$, $\alpha = \pi/2$, and multiples thereof. 
	In the following calculations, we set $\alpha=\pi/7$. 
	We always impose open boundary condition in the calculations because of the quasiperiodicity.
	In order to simulate many samples, 
	we randomize the arbitrary phases $\{\phi_i\}$ in the cosines within $(0, 2\pi]$.
	These systems will be equivalent by an overall translation in the thermodynamic limit.

	\section{Phase diagram and mobility edge}\label{sec_PD_ME}
	We first study the phase diagrams of these models spanned by the eigenenergy $E$ and the QP strength $V$.
	Here, we diagonalize the Hamiltonians in cubic systems with linear size $L$ 
	and obtain the eigenenergies $\{E_i\}$ and the normalized eigenfunctions $\{\psi_i(\bm r)\}$. 
	The participation ratio for the eigenfunction $\psi_i(\bm r)$ is defined as
	\begin{align}
		{\rm PR}(E_i)\equiv\frac{1}{\sum_{\bm r}|\psi_i(\bm r)|^4}. \label{PR}
	\end{align}
	The delocalization/localization states can be characterized by the normalized participation ratio \cite{Li17,Li20,Roy21} in the logarithmic form,
	which is defined as 
	\begin{align}
		\eta(E_i)\equiv\frac{\ln {\rm PR}(E_i)}{\ln L}. \label{D2}
	\end{align} 
	When the wavefunction is well delocalized, ${\rm PR}\sim L^3$ and $\eta\approx3$.
	On the contrary, when the wavefunction is localized, 
	${\rm PR}\sim \xi^3$ with $\xi$ the localization length, and $\eta\approx0$. 
	%\st{As $L\rightarrow\infty$, ${\rm NPR}$ is the fractal dimension [3,62,63].} 
	$\eta$ is plotted in the plane spanned by $E$ and $V$ (Fig. \ref{fig_D2})
	for the three quasiperiodic models introduced in Section \ref{sec_model}.
	As seen in Fig. \ref{fig_D2}, the values of $\eta$ change rapidly,
	indicating the existence of mobility edges which fluctuate significantly as we vary QP strength $V$. 
	
	From now we focus on the Fermi energy level $E=0$ and analyze the critical behavior.
	In order to characterize the delocalization-localization transitions, the localization length
	and conductance are calculated by the transfer matrix method.
	Then the critical exponents are extracted by the polynomial fitting method.
	We note that the fitting is sensitive to the statistical error estimates of the fitting data,
	and the incorrect  error estimates result in incorrect evaluation of critical exponents.
	Therefore we should be careful of the statistical error estimates of the localization length 
	and conductance in quasiperiodic systems.
	We argue that there is a correlation between samples for localization length and conductance resulting from the QP,
	which result in incorrect error estimates of them.
	However, the correlation can be controlled by the limited number of samples
	(see the detailed analysis in Appendices \ref{Appendix_auto_g} and \ref{Appendix_auto_lambda}).
	In this paper, we take conductance as an example to study the critical behavior.
	We think the localization length analysis also gives a similar result if the error estimates are correctly done.
	
	In the next section, we will introduce the calculation of conductance
	and perform the finite size scaling analysis to extract the critical parameters, such as critical exponent $\nu$ and
	critical quasiperiodic potential strength $V_c$.
	
	\section{Finite size scaling of the conductance}\label{sec_FSS}
	The zero temperature two-terminal conductance in units of $e^2/h$ is given by 
	\begin{align}
		g={\rm tr}(tt^{\dagger})
	\end{align} 
	with $t$ the transmission matrix.
	The transmission matrix $t$ can be calculated by the transfer matrix method \cite{Pendry92}.
	The two-terminal conductance is calculated along $z$ direction with ideal lead in the two side of the cubic system ($L_x=L_y=L_z=L$).
	%In order to keep the quasi-periodicity,  the open boundary condition is imposed on the transverse direction
	%($x, y$ direction) in the following calculation.
	
	%The mean conductance $\langle g\rangle\equiv \frac{1}{N}\sum_i^{N}g_i$ has been averaged by various samples
	%$N$ with different initial
	%phase $\{\phi_i\}$. Then the standard error of $\langle g\rangle$ has been estimated as following,

\begin{figure}[!tb]
		\centering
		\includegraphics[width=1\linewidth]{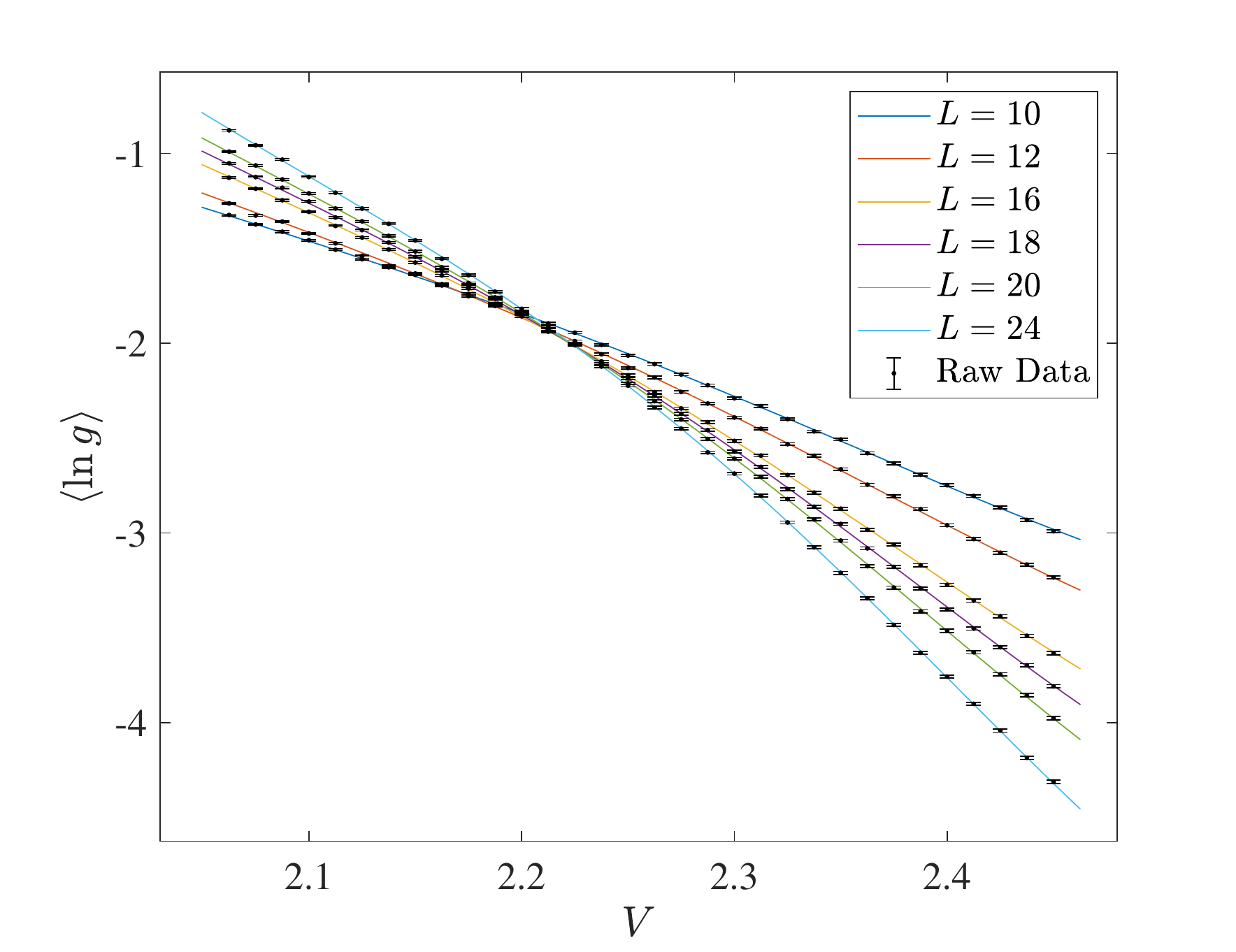}
		\includegraphics[width=1\linewidth]{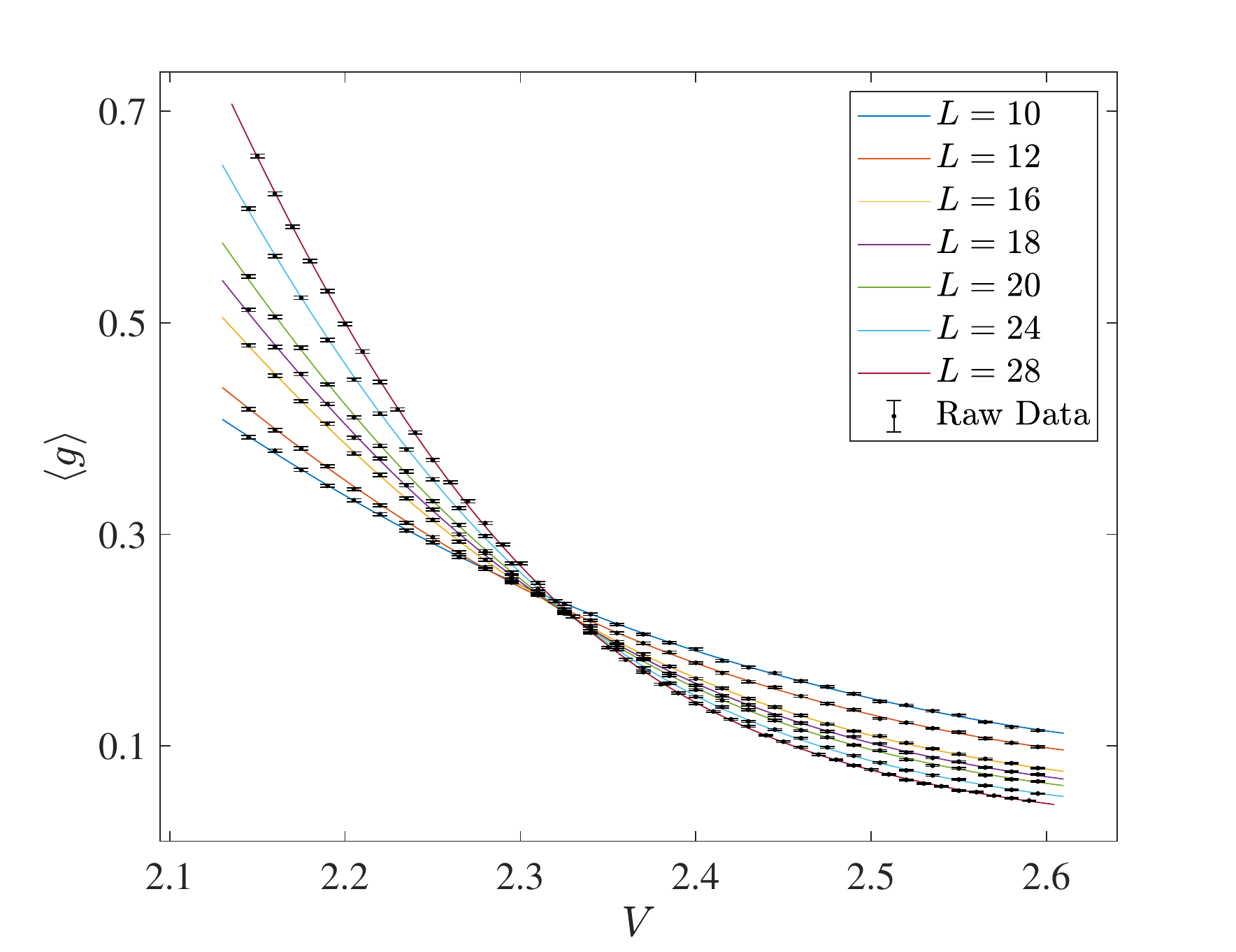}
		\includegraphics[width=1\linewidth]{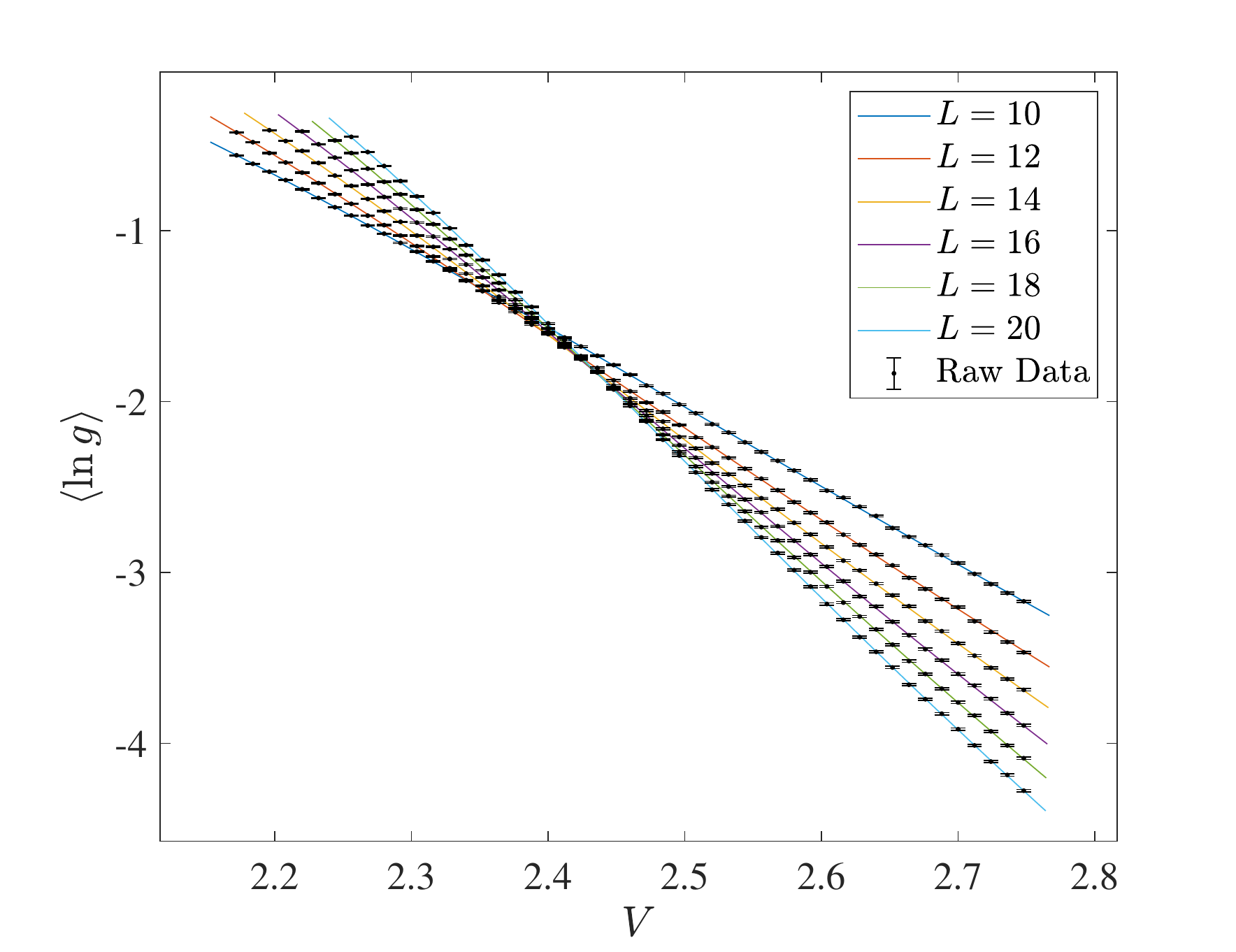}
		\caption{The two-terminal conductance as a function of the strength of quasiperiodic
			potential $V$ for Anderson, Peierls phase, and Ando models (from top to bottom).
			The black points with the error bars are the raw data and lines with different colors  
			are the polynomial fitting results with expansion order ($m_1$, $n_1$, $m_1$, $n_1$) 
			= (3, 3, 0, 1). }
		\label{scaling_g}
	\end{figure}

	%Under the control of the correlation (see Appendix \ref{Appendix_auto_g}), 
	The mean conductance $\langle g\rangle$  is averaged over $50000$ samples 
	for the AM and PPM, $10^5$ samples for the Ando model with random phase $\{\phi_i\}$.
	We note that $g(\{\phi_i\})$ and $g(\{\phi_i+\delta_{\phi_i}\})$ are correlated for small $\delta_{\phi_i}$, but
	uncorrelated when $\delta_{\phi_i}$ is large enough (see Appendix \ref{Appendix_auto_g}), 
	just like the magnetoconductance in mesoscopic systems \cite{Umbach84,Stone85,Webb85,Lee85,Daimon22}.
	The behavior of the mean conductance $\langle g\rangle$ or logarithmic average $\langle \ln g \rangle$  
	with error bar are shown in Fig.~\ref{scaling_g} for various $V$ and $L$.
	The scale invariant point of conductance indicates the critical point $V_c$ of the AT
	driven by QP. With $V<V_c$ ($V>V_c$), the conductance increases (decreases) with system
	size $L$, indicating it is in the delocalized (localized) phase.
	To extract the universal critical parameters near the critical point, %such as critical exponent,
	the polynomial fitting method based on the finite size scaling theory is introduced as follows.
	
	\begin{figure*}[bt]
		\centering
		\subfigure[Three quasiperiodic models]{
			\begin{minipage}[t]{0.5\linewidth}
				\centering
				\includegraphics[width=1\linewidth]{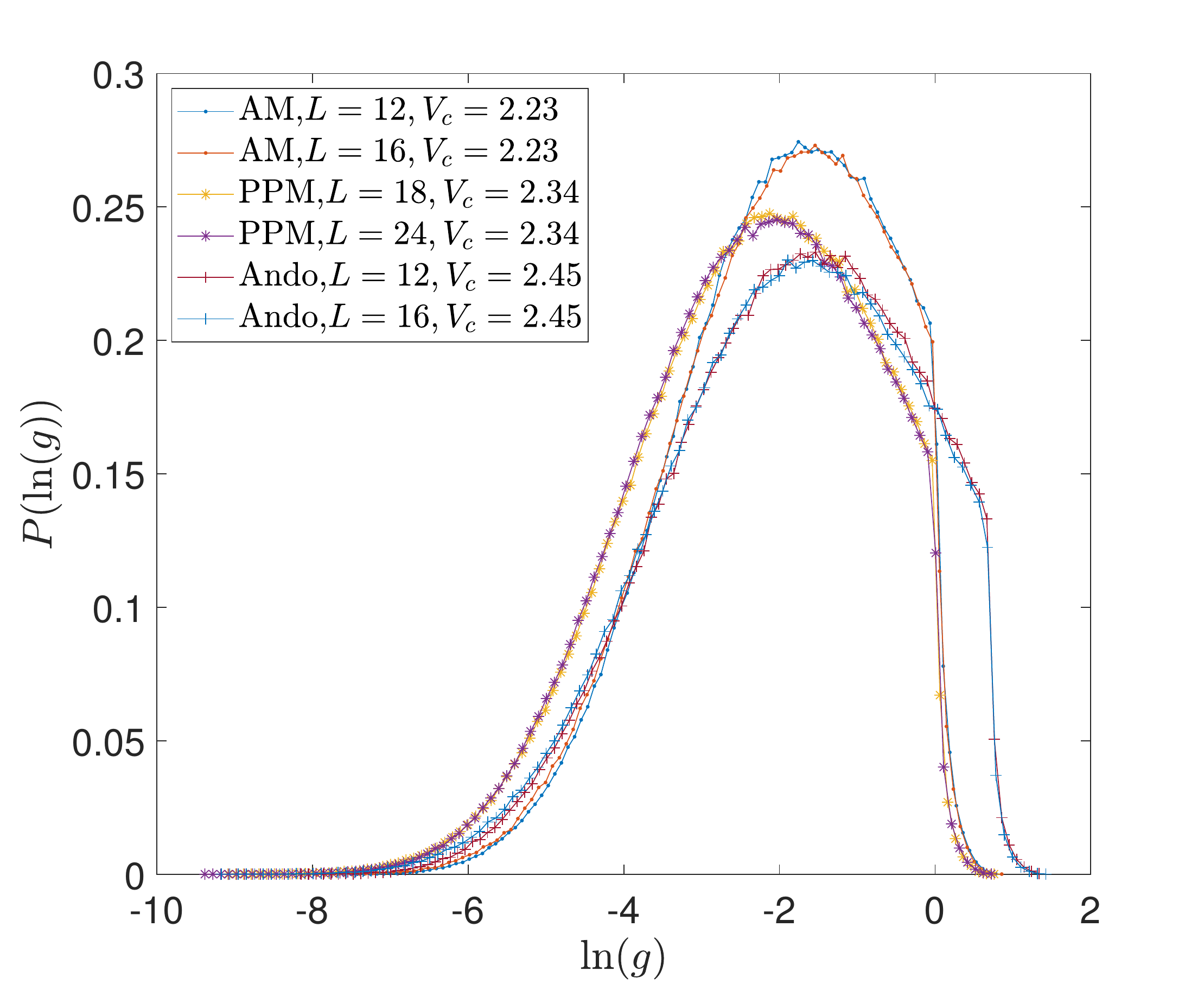}
				%{AAH-Ando-PPM_G_Comparison_CCD }
				%\caption{fig1}
			\end{minipage}%
		}%
		\subfigure[Anderson model]{
			\begin{minipage}[t]{0.5\linewidth}
				\centering
				\includegraphics[width=1\linewidth]{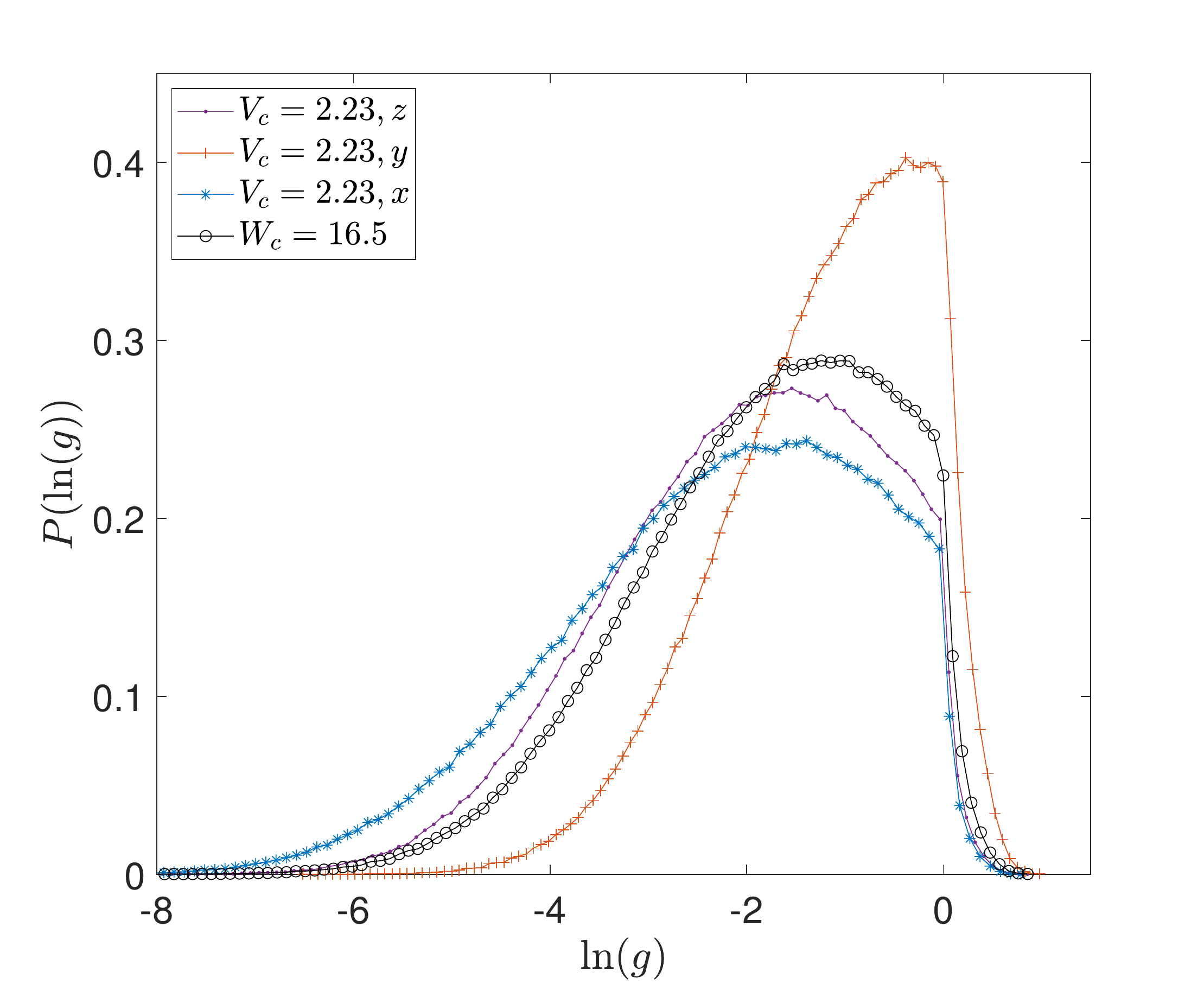}
				%{ AAH_3D_couple_G_CCD_zxy_WithAM_L16 }
				%\caption{fig1}
			\end{minipage}%
		}%	
		\caption{(a) Critical conductance distributions (CCDs)  along $z$ direction
			for the Anderson model (AM), Peierls phase model (PPM), and Ando model at critical quasiperiodic potential $V_c$ estimated in TABLE \ref{table_nu}. 
			(b) The CCDs of quasiperiodic AM along $x$, $y$, and $z$ directions. For comparison, 
			the CCD of disordered AM at critical disorder $W_c$ is also shown in circles. 
			The two-terminal conductance $g$ is calculated in cubic system with linear size $L$ 
			and $L=16$ for (b). 
			Data for each distribution comes from $10^6$ samples. }
		\label{fig:gcomparison}
	\end{figure*}
	
	The single parameter scaling theory \cite{Abrahams79} assumes that the dimensionless physical quantity $\Gamma$, 
	such as the normalized localization length and the two-terminal conductance,
	obeys the universal scaling function,
	\begin{align}
		\Gamma=f(\phi_1L^{1/\nu}), \label{SingleParameterScaling}
	\end{align}
	where $\phi_1$ is the relevant scaling variable, $L$ is the system size, and $\nu$ is the critical exponent
	describing the power-law divergence of the length scales near the critical point.
	For the actual fitting, we often need to take into account the irrelevant scaling variables 
	because of the corrections to the single parameter scaling \cite{Slevin99,Slevin14}.
	Here we just consider the least-irrelevant scaling variable $\phi_2$, 
	then the universal scaling function can be written as
	\begin{align}
		\Gamma=F(\phi_1L^{1/\nu}, \phi_2L^{-y}), \label{F}
	\end{align}
	where $-y$ is the scaling dimension for the irrelevant scaling variable.
	The term $\phi_2L^{-y}$ will decrease to zero in power-law with $L$ increasing and the scaling function 
	returns to the single parameter scaling form in Eq. (\ref{SingleParameterScaling}) in the thermodynamic limit.
	The relevant or irrelevant scaling variables depend on the system parameter driving the AT.
	In this paper, QP strength $V$ is this system parameter, 
	and a normalized distance of $V$ from the critical point $V_c$ 	is define as $w\equiv (V-V_c)/V_c$. 
	Then scaling variables $\phi_1$ and $\phi_2$ are functions of $w$ with 
	$\phi_1(w=0)=0$ by definition. Considering the possible non-linearity of $w$ near the critical point, 
	$\phi_1$ and $\phi_2$  are expanded in powers of small $w$ when $V$ is sufficiently close 
	to $V_c$;  
	\begin{equation}
		\phi_{i}(w) \equiv \sum^{m_i}_{j=0} b_{i,j} w^{j}	 \label{uiw}
	\end{equation}
	with $i=1,2$, $b_{1,0}=0$. 
	When $V$ is sufficiently close to $V_c$ and the values of $\Gamma$ are controlled in a finite range, 
	the universal scaling function $F$ in Eq. (\ref{F}) can also be expanded in powers of its small arguments,   
	\begin{equation}
		F = \sum^{n_1}_{j_1=0} \sum^{n_2}_{j_2=0} a_{j_1,j_2} (\phi_1L^{1/\nu})^{j_1} (\phi_2L^{-y})^{j_2}. \label{Fexp}
	\end{equation}
	To reduce the redundant two degrees of freedom in Eq. (\ref{Fexp}), 
	we set $a_{0,1}=a_{1,0}=1$.
	We note that the assumption here is a posteriori justified by the numerical analysis.
	%	\textcolor{red}{TO: it is the same thing practically, but it might be better to say
	%		$a_{1,0}=a_{0,1}=1$ instead of saying $b_{1,1}=b_{2,0}=1$.
	%		This is because $F$ is a universal function.}

	Given $(m_1,n_1,m_2,n_2)$ in Eqs.~(\ref{uiw}) and (\ref{Fexp}), $F$ is a  
	finite-order of polynomial of $L$ and $w \equiv (V-V_c)/V_c$. Numerical data of $\Gamma$ 
	for different $L$ and $V$ are fitted by the polynomial  
	with fitting parameters $V_c$, $\nu$, $y$, $a_{i,j}$ and $b_{i,j}$.
	We minimize $\chi^2$ in terms of the fitting parameters,    
	\begin{equation}
		\chi^2 \equiv \sum^{N_D}_{k=1} \frac{(\Gamma_{k}-F_{k})^2}{\sigma^2_{\Gamma_{k}}}.	\label{key}
	\end{equation}
	Here $k$ counts the data points ($k=1,\cdots, N_D$), and each data point 
	is specified by $L$ and $V$; $k=(L,V)$. $\Gamma_{k}$ and $\sigma_{\Gamma_{k}}$ 
	are a mean value of dimensionless physical quantity and its standard deviation at $k=(L,V)$, respectively, 
	while $F_{k}$ is the fitting value from the polynomial $F$ at $k=(L,V)$. 
	$F_k$ depends 
	on the fitting parameters and $\chi^2$ is minimized in terms of them. 
	The minimization is carried out for several different choices of $(m_1,n_1,m_2,n_2)$. 
	The quality of the fitting is characterized by the goodness of fit in statistics \cite{Bevington03,Press92}.
	The 95$\%$ confidence intervals for the fitting results are determined by 1000 sets of 
	$N_D$ synthetic data points,  
	which are statistically generated from the fitting value $F_k$ with
	the same standard deviation of $\Gamma_k$ at each point $k$.
	
	Here the dimensionless physical quantity conductance [$\langle g\rangle$ or $\langle \ln g\rangle$]
	will be fitted by the polynomial fitting method.
	The fitting results with the goodness of fit greater than 0.1 are shown in Table~\ref{table_nu}.
	$V_c$ and $\nu$ are shown to be robust against the change of the 
	expansion orders and various system sizes in Table \ref{table_nu}.
	The critical exponents $\nu$ are estimated as $\nu=1.613~[1.587, 1.639]$ for the AM,
	$\nu=1.45~[1.42, 1.49]$ for the PPM, and $\nu=1.37~[1.35, 1.39]$ for 
	the Ando model, which are consistent with the critical exponents of ATs  driven by random potential,
	i.e., $\nu\approx 1.57$ \cite{Slevin14} for 3D orthogonal class, $\nu\approx1.43$ \cite{Slevin99} for 3D
	unitary class, 
	and $\nu\approx 1.36$ \cite{Asada05} for 3D symplectic class, respectively.
	We also find that the critical QP strength $V_c$ increases for the PPM and Ando model compared with that for the AM,
	which is similar to that in disordered systems.
	This increase of $V_c$ by the magnetic field (Peierls phase model) and spin-orbit interaction (Ando model) is consistent with the weak localization \cite{Lee85RMP}.
	In conclusion, the universality classes of ATs driven by QP are similar
	to those of ATs driven by the random potential in the 3D Wigner-Dyson symmetry classes \cite{Devakul17}.
	
	\section{Critical conductance distribution}\label{sec_CCD}
	\begin{figure}[bt]
		\centering
		\includegraphics[width=1.1\linewidth]{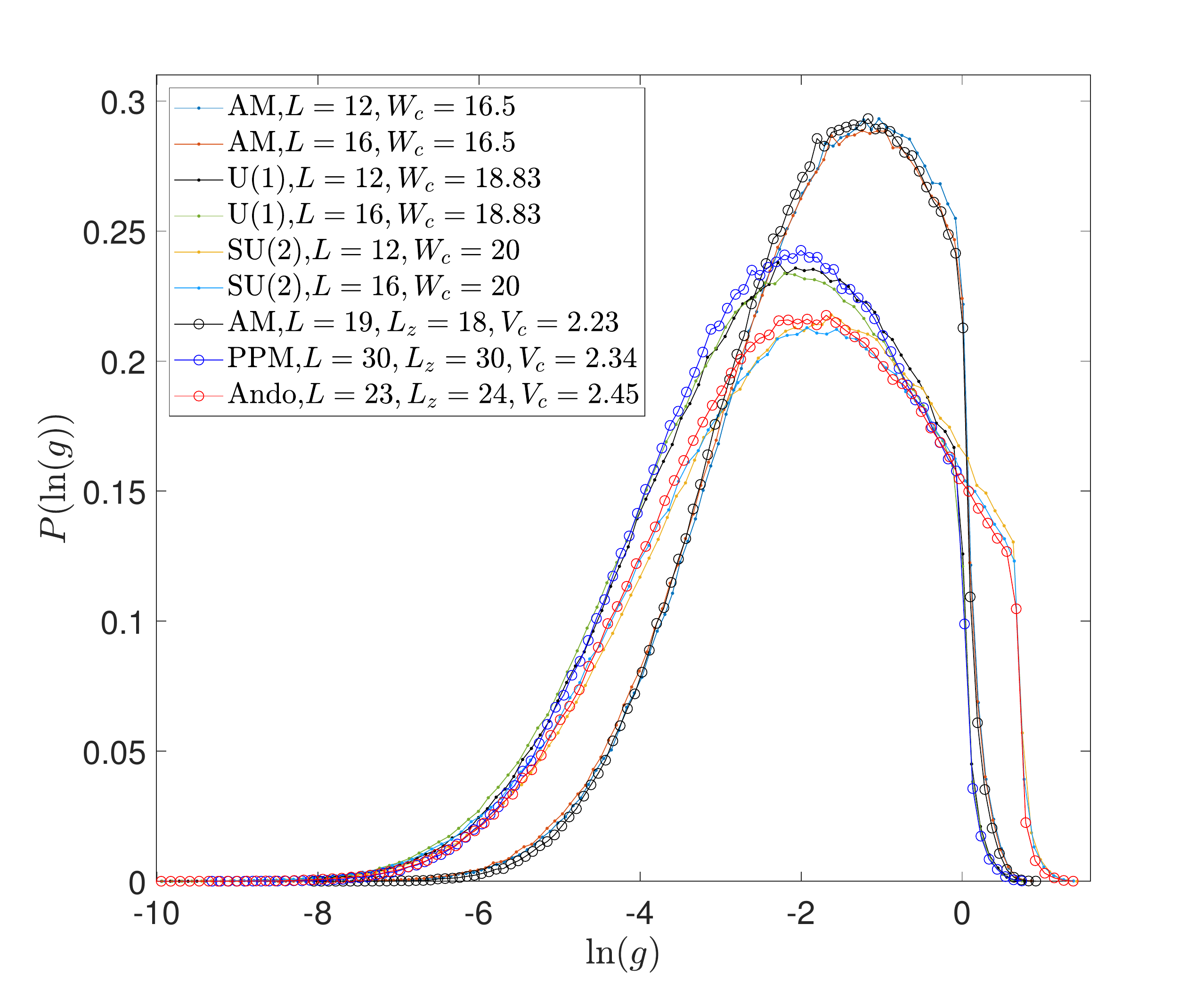}
		%{WignerDysonClass_quasi_tuned_VS_disorder}
		\caption{Comparison of the critical conductance distributions (CCDs) between quasiperiodic and disordered systems.
			The lines with points stand for the CCDs of disordered systems at critical disorder $W_c$,
			i.e., Anderson model (AM), U(1) model, and SU(2) model in cubic systems with linear size $L$.
			The lines with circles are the CCDs for quasiperiodic systems at critical quasiperiodic potential $V_c$, 
			i.e., AM, Peierls phase model (PPM) and Ando model, 
			with tuned geometry (cross-section linear size $L$ and transmission length $L_z$).
			Data for each distribution comes from $10^6$ samples.}
		\label{fig:AAH_3D_couple_G_CCD_zxy}
	\end{figure}
	
	To reinforce our conclusion that the universality classes in 3D quasiperiodic systems
		are the same as the cases in disordered systems,
	we have also studied the critical conductance distributions (CCDs)
	and compared them with those in disordered systems. 
	
	At the critical point, the correlation length will diverge and the conductance distribution
	will not depend on the system size $L$, which has been shown in Fig. \ref{fig:gcomparison} (a).
	This also indicates that the estimates of critical points are correct.
	
	The CCDs along $x$ and $y$ directions are also calculated for the 3D AM.
	We notice that the CCDs along $x$, $y$, and $z$ directions are significantly different from each other
	[Fig. \ref{fig:gcomparison} (b)], which indicates the system is anisotropic because of the QP we use.
	We also find that the average conductance along $x$, $y$, and $z$ directions, i.e., 
	$\langle g_x \rangle \approx 0.24$, 
	$\langle g_y \rangle \approx 0.47$,
	$\langle g_z \rangle \approx 0.26$,
	and the average conductance of AM with random potential 
	$\langle g \rangle \approx 0.3$
	satisfy the relationship
	$\langle g_x \rangle \langle g_y \rangle \langle g_z \rangle \approx \langle g \rangle^3$.
	We note that this relation is similar to that for the quasi-1D localization length
	\cite{Zambetaki96}.
	
	To test the universality classes of the ATs driven by QP,
	we compare the CCDs between disordered and quasiperiodic systems.
	Besides the symmetry and dimension, CCD also depends on the topology 
	\cite{Slevin00} (i.e., the boundary conditions) and effective geometry \cite{Luo18QMCT} 
	(i.e., the ratio of the linear sizes between transverse direction and transmission 
	direction, or the mean value of conductance).
	In order to compare the CCDs, we need to keep the symmetry, dimension, boundary condition, and 
	effective geometry the same.
	
	For the disordered systems, we impose the open boundary condition in the transverse direction
	and calculate the conductance in the cubic geometry.
	Anderson model, U(1) model \cite{Slevin16}, and SU(2) model \cite{Asada05} with random potential
	are chosen as the examples of  Wigner-Dyson symmetry classes because of their isotropic property.
	We add the random potential with a box distribution $[-W/2, W/2]$ to these models.
	The critical disorder $W_c$ is believed to be independent of the boundary conditions \cite{Slevin00}
	so here we just use $W_c$ estimated in systems with periodic boundary condition for the CCD calculations.

	The effective geometry in quasiperiodic systems is different from the real geometry
	because of the anisotropic conductance caused by the QP.
	Following the idea in Ref. \onlinecite{Luo18QMCT}, 
	we can fine-tune the geometry ratio of these quasiperiodic models;
	$\rho\equiv L/L_z$ with $L$ the size in the transverse direction and $L_z$ the transmission length,
	so that the mean value of conductance for quasiperiodic and isotropic disordered systems is almost the same.
	The ratios are set as  $\rho=19/18$, $1$, and $23/24$, for the AM, PPM, and Ando model, respectively.
	Fig. \ref{fig:AAH_3D_couple_G_CCD_zxy} shows that CCDs of quasiperiodic and disordered systems are almost 
	similar, which again indicates that the universality classes of the ATs in quasiperiodic and disordered systems are the same.
	
	\begin{figure}[bt]
		\centering
		\includegraphics[width=1.1\linewidth]{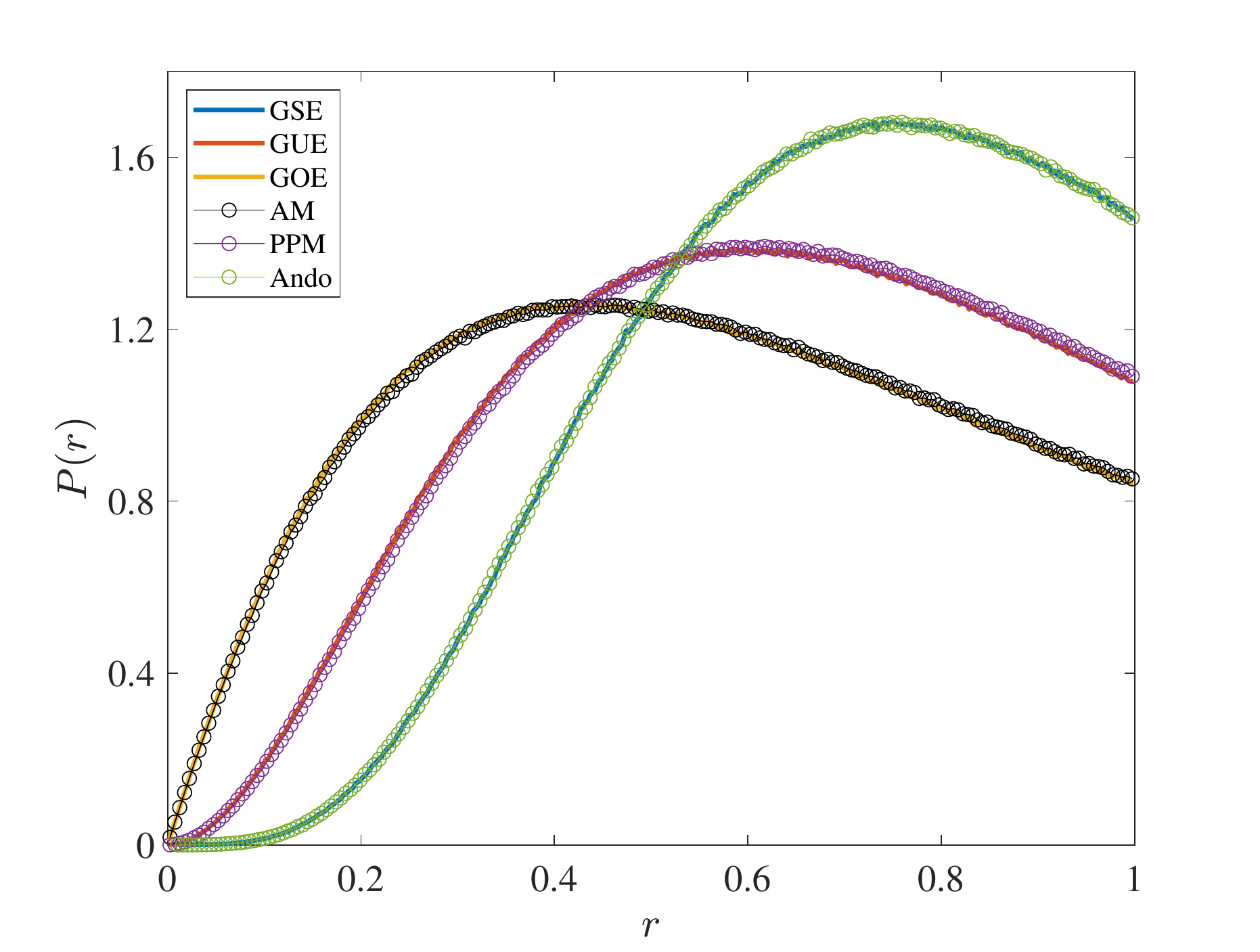}
		%{Pr_GOE_GUE_GSE_quasi}
		\caption{Level spacing ratio distribution $P(r)$.
			The solid lines stand for the distributions of the random matrix in 
			the Gaussian orthogonal ensemble (GOE), Gaussian unitary ensemble (GUE), 
			and Gaussian symplectic ensemble (GSE).
			The circles are the distributions in the delocalized states of the Anderson model (AM), 
			Peierls phase model (PPM), and Ando model.  }
		\label{fig_pr}
	\end{figure}
	
	\begin{figure*}[bt]
		\centering
		% subfigures
		\subfigure[Anderson model ]{
			\begin{minipage}[t]{0.333\linewidth}
				\centering
				\includegraphics[width=1\linewidth]{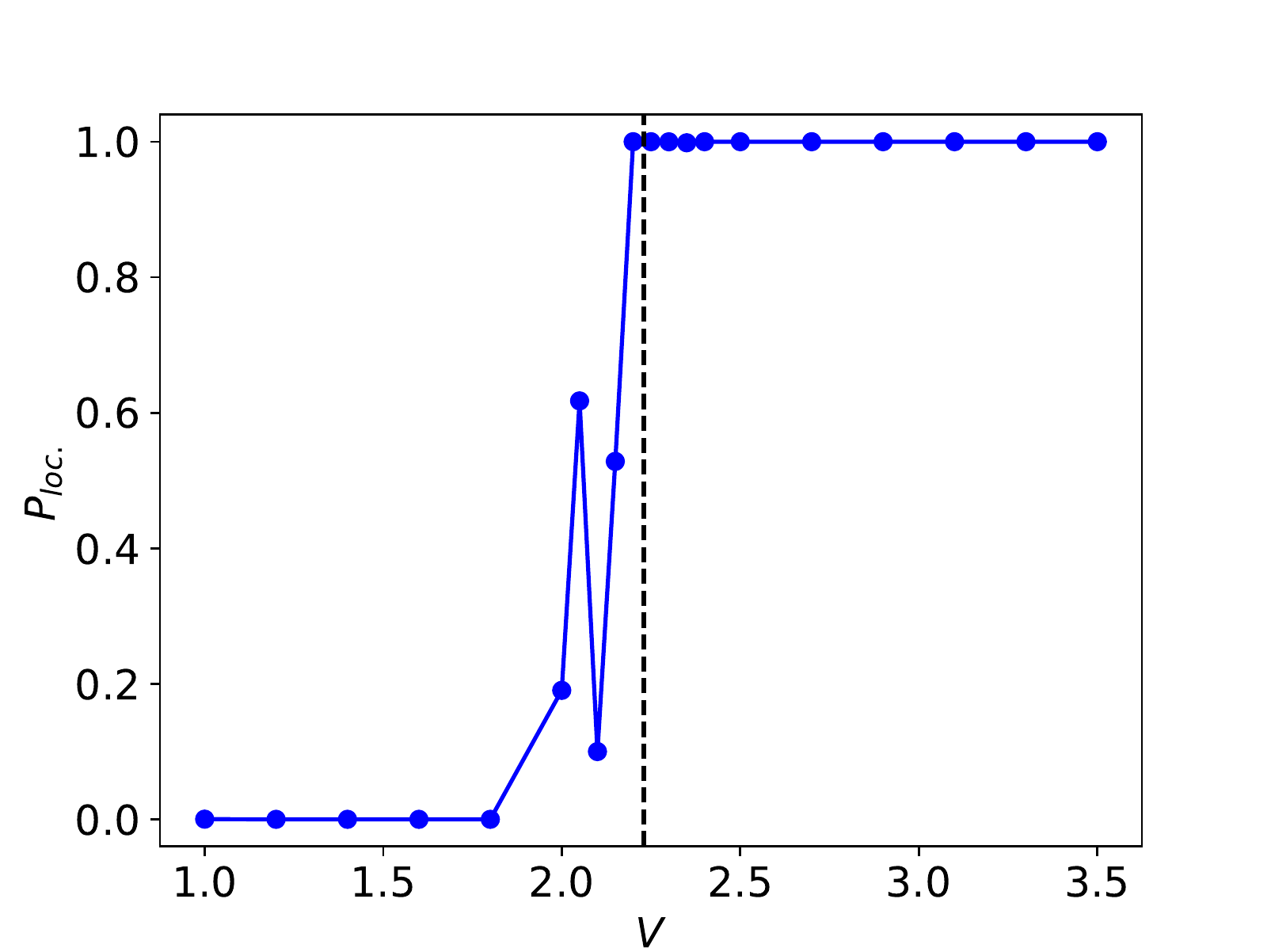}
				%{P_AAH}
				%\caption{fig1}
			\end{minipage}%
		}%
		\subfigure[Peierls phase model ]{
			\begin{minipage}[t]{0.333\linewidth}
				\centering
				\includegraphics[width=1\linewidth]{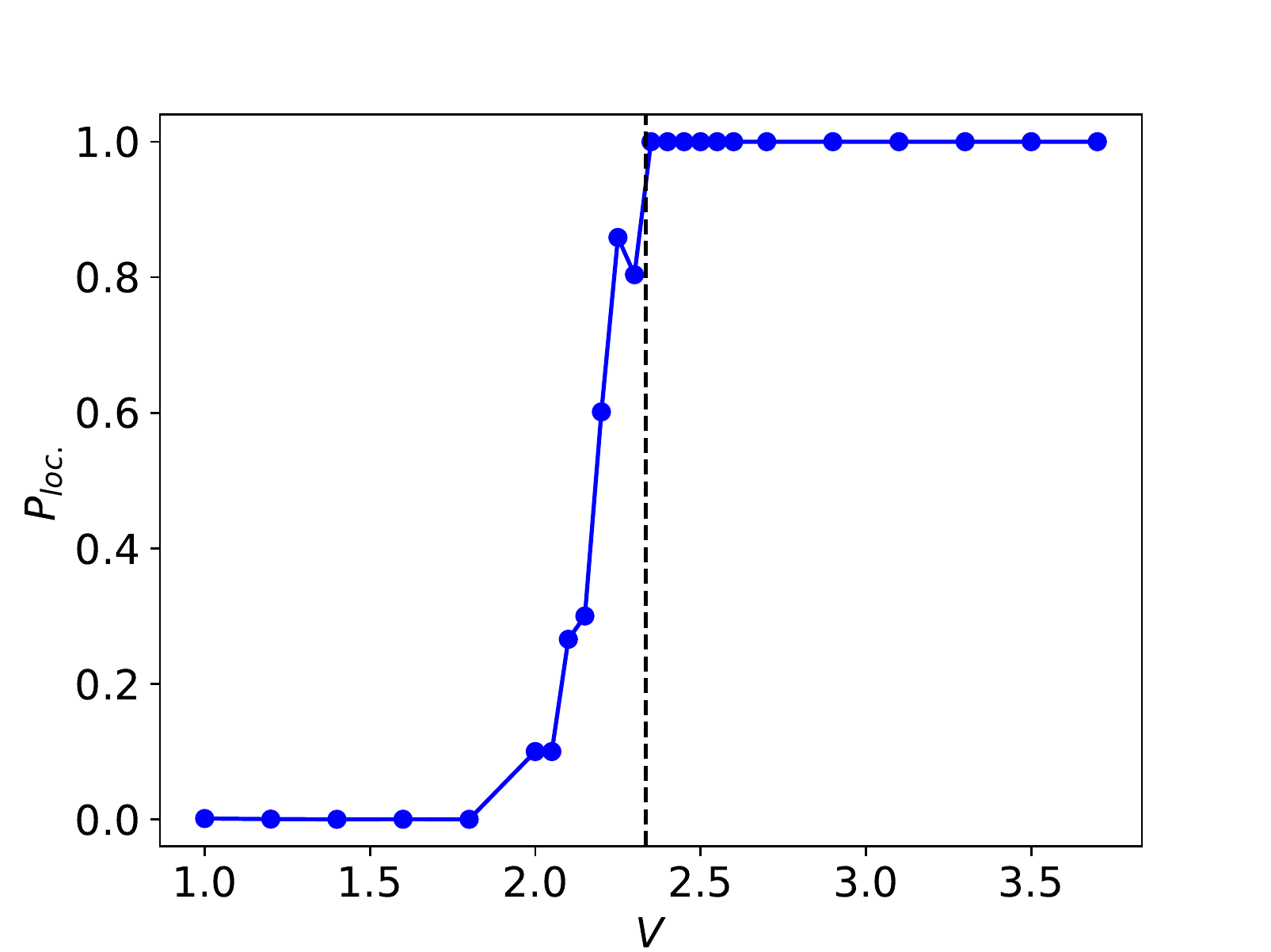}
				%{P_PPM}
				%\caption{fig1}
			\end{minipage}%
		}%
		\subfigure[Ando model ]{
			\begin{minipage}[t]{0.333\linewidth}
				\centering
				\includegraphics[width=1\linewidth]{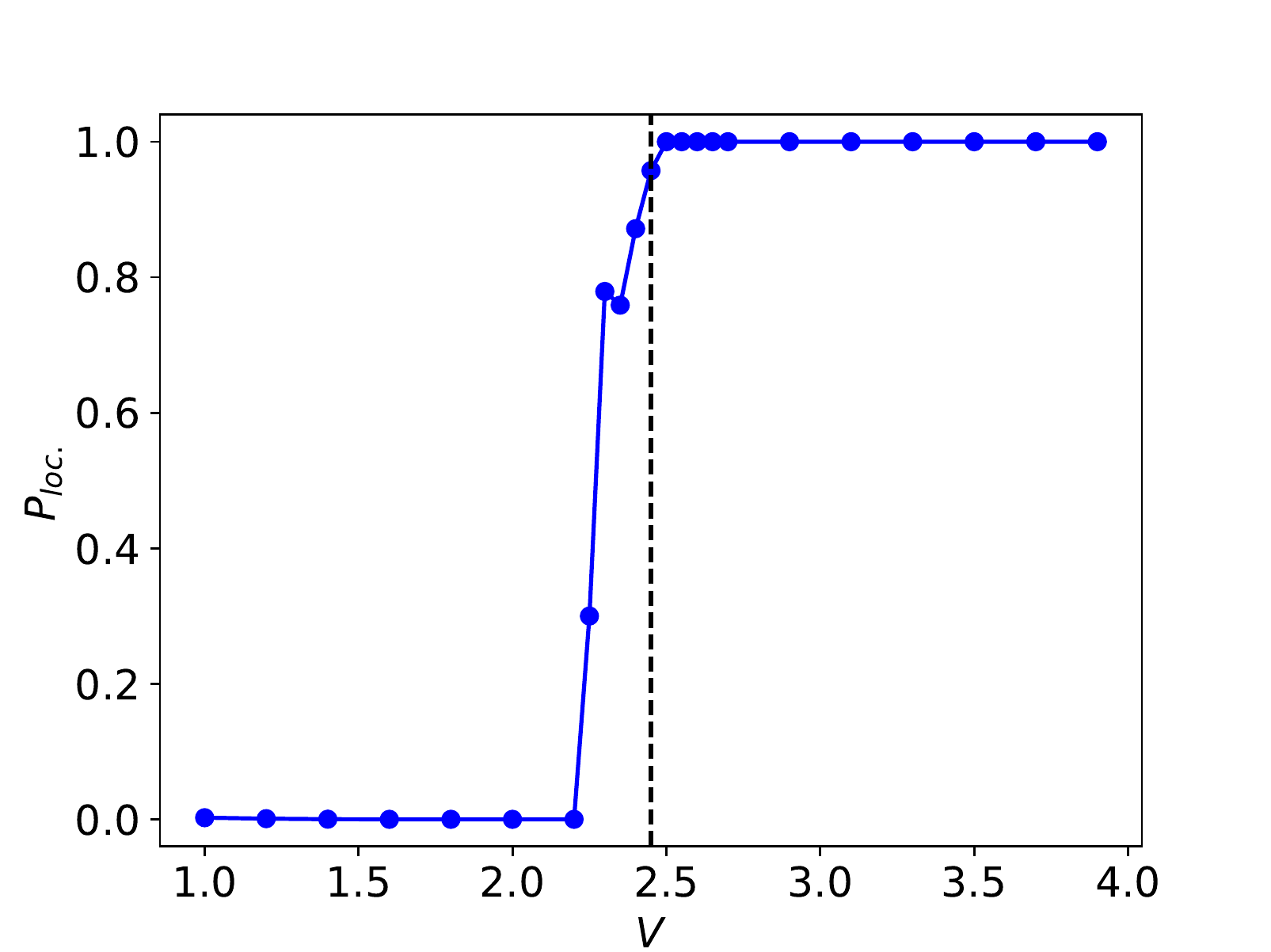}
				%{P_Ando}
				%\caption{fig1}
			\end{minipage}%
		}%
		\caption{The probability of localization $P_{loc.}$ for wavefunctions in quasiperiodic systems
			as a function of quasiperiodic potential strength 
			$V$ predicted by a convolutional neural network trained by
			the localized/delocalized wavefunctions in the disordered Anderson model.
			The critical points $V_c$ estimated in Table \ref{table_nu} are indicated
			by the vertical dashed lines as a guide to the eye.
			The probability of each data point is averaged by 10 samples.
			One wavefunction is selected near $E=0$ in a cubic system with linear size $L=42$ for each sample
			with random $\{\phi_{i}\}$.
		}
		\label{fig_ccn_predict}
	\end{figure*}
	
	\section{Level spacing ratio distribution}\label{sec_Pr}
	The level spacing ratio has been introduced to characterize the energy level repulsion in Ref. \onlinecite{Oganesyan07}.
	Let $\{E_i\}$ be a series of energy levels indexed in ascending order 
	and $s_i=E_i-E_{i-1}$ is the nearest-neighbor spacing.
	Then the level spacing ratio is defined as
	\begin{align}
		r_i\equiv\frac{\min\{s_i,s_{i-1}\}}{\max\{s_i,s_{i-1}\}}.
	\end{align}
	According to the Wigner-Dyson random matrix theory \cite{Wigner51,Dyson62,Dyson62TFW}, 
	the level spacing ratio distribution $P(r)$ follows the Wigner-like
	surmises for all the three random matrix ensembles \cite{Atas13},
	i.e., Gaussian orthogonal ensemble (GOE), Gaussian unitary ensemble (GUE), and Gaussian symplectic ensemble (GSE).
	%In disordered system, $P(r)$ in the metal phase of the corresponding symmetry system follows the Wigner-like surmises.
	$P(r)$ in the delocalized states of the 1D AAH model doesn't follow the distribution
	of GOE \cite{Li16,Deng17}.
	On the other hand, $P(r)$ for the non-interacting Hamiltonians defined on the 2D
	quasiperiodic Ammann-Beenker tiling \cite{Grimm21}, 2D Moir\'e lattice \cite{Huang19}, 
	and 3D generalized AAH model \cite{Devakul17} show the distribution expected from GOE.
	
	Here we study the level spacing ratio distribution in the three quasiperiodic models.
	We diagonalize the Hamiltonian in a cubic system with linear size $L=10$.
	We set $V=0.5$ for the three models and take half of the eigenenergy near $E=0$ 
	so that all of the states are delocalized (see Fig. \ref{fig_D2}). 
	$6\times10^4$ samples with random phase $\{\phi_i\}$ have been simulated for the distribution.
	We observe that $P(r)$ in the delocalized states of the AM, PPM, and Ando model 
	well coincide with the distributions of GOE, GUE, and GSE, respectively (Fig. \ref{fig_pr}).
	
	\section{Convolutional neural network study of the wavefunction}\label{sec_CNN}
	Deep neural networks have been trained to predict the delocalization/localization probability
	and topological/non-topological probability
	of the wavefunctions in the 2D \cite{Ohtsuki16} and 3D \cite{Ohtsuki17,Mano17} disordered systems.
	Here we use a convolutional neural network (CNN) whose parameters have been trained by the localized/delocalized 
	wavefunctions in the 3D disordered Anderson model.
	The details of the network structure is found in Ref. \cite{Ohtsuki20}.
	We input the wavefunctions of quasiperiodic systems at $E=0$ to this CNN,
	and let it determines the localization/delocalization probability.
	The predicted critical points (Fig. \ref{fig_ccn_predict}) well coincide with those estimated by conductance,
	which means that the localization behavior in 3D QP systems is similar to the ones in random systems.
	
	\section{Summary and concluding remarks }\label{sec_summary}
	In this paper, we have used three models in the 3D Wigner-Dyson symmetry classes to study
	the critical behavior of the ATs driven by QP.
	The two-terminal conductance and localization length have been calculated by the transfer matrix method.
	We find there is a correlation between samples, which results in the wrong estimation of their statistical errors.
	By carefully taking account of the statistical independence of samples,
	the critical exponents $\nu$ and critical points $V_c$
	have been estimated by the finite size scaling analysis of two-terminal conductance.
	The estimated critical exponents $\nu$ coincide with those of the ATs driven by random potential with the same symmetries.
	Our estimations of $\nu$ for the AM and Ando model are consistent with those reported in Ref. \onlinecite{Devakul17,Sutradhar19}.
	Moreover, the critical conductance distribution and the level spacing distribution have also been studied.
	We also find that a convolutional neural network trained by the localized/delocalized
	wavefunctions in a disordered system can determine whether the wavefunctions in quasiperiodic systems
	are localized or not.
	These numerical results strongly support that the universality classes of ATs driven
	by QP and random potential are the same in the 3D Wigner-Dyson symmetry classes.
	
	%For 1D AAH model, the Lyapunov exponent calculated by transfer matrix method 
	%will become nonzero abruptly when increasing $V$ through $V_c=1$,
	%where the correlation length will diverge as $\xi\propto 1/|\ln V|$ \cite{Deng17}.
	%Hence the critical exponent for the correlation length is exactly $\nu=1$.
	%For 2D case, the calculation of conductance or localization length is not smooth as a function of $V$, 
	%and the critical exponent can't be estimated precisely.
	
	Considering the recent works about the universality classes of the Anderson transitions
	in non-Hermitian disordered systems \cite{Huang20,Luo21,Luo21TM,Luo22},
	we can also extend the present work to check the universality classes in the non-Hermitian systems with QP,
	which is an interesting topic left for the future.
	%And we leave it for the future work.
	
	%	\bigskip
	%	\bigskip
	
	\begin{acknowledgments}
		X. L. was supported by National Natural Science Foundation of China (Grants No.12105253 and 11934020) and the Project supported by CAEP Foundation (Grant No. CX20210035).
		T. O. was supported by JSPS KAKENHI Grants No. 19H00658 and 22H05114. 
		%R. S. was supported by the National Basic Research Programs of China (No. 2019YFA0308401) 
		%and by National Natural Science Foundation of China (No.11674011 and No. 12074008). 
	\end{acknowledgments}
	%\clearpage
	
	\appendix
	
	\begin{figure*}[tb]
		\centering
		\subfigure{
			\begin{minipage}[t]{0.25\linewidth}
				\centering
				\includegraphics[width=1\linewidth]{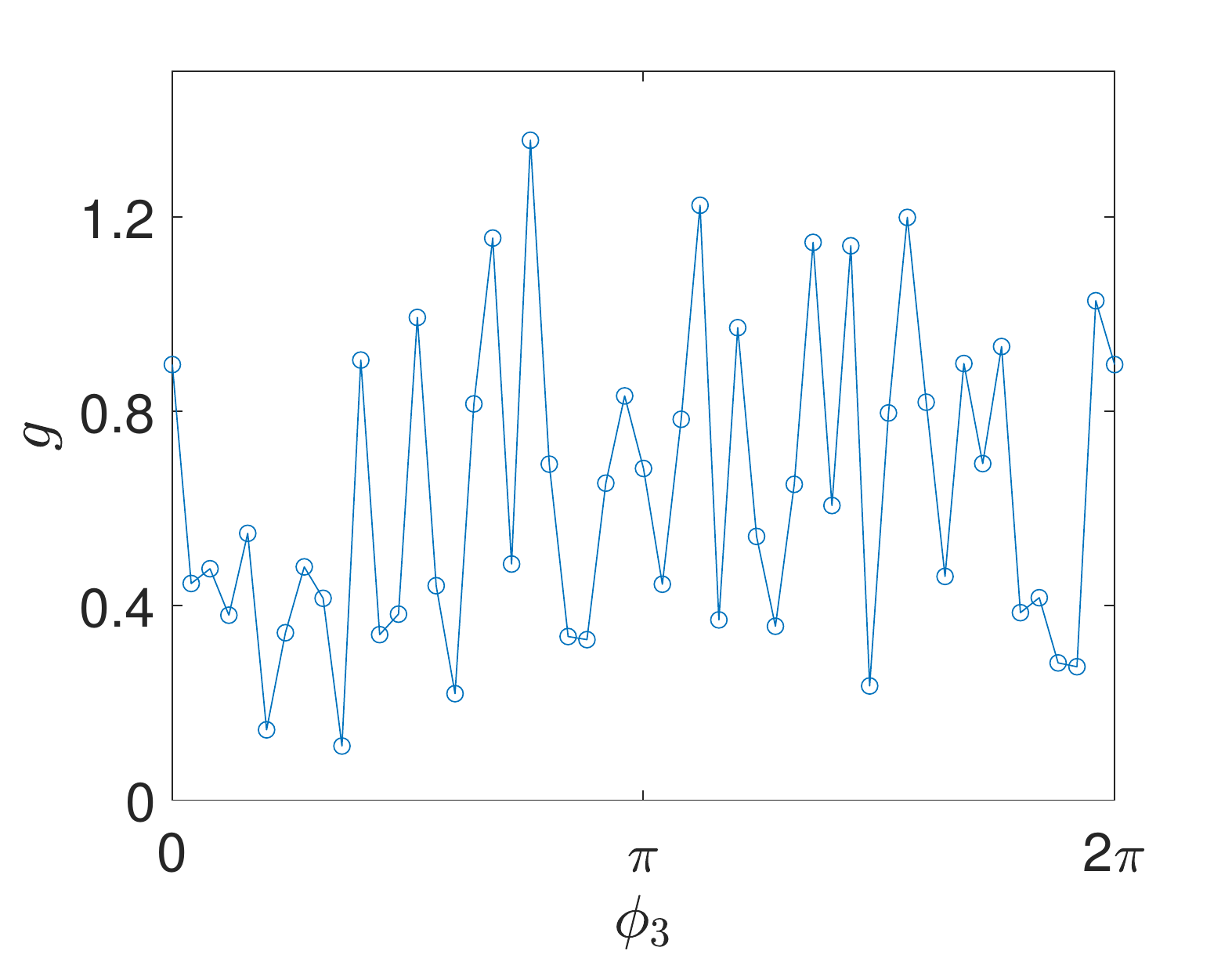}
				% {AAH_3D_couple_L20_V2_phi1_05_phi2_05_g.eps}
				%\caption{fig1}
			\end{minipage}%
		}%
		\subfigure{
			\begin{minipage}[t]{0.25\linewidth}
				\centering
				\includegraphics[width=1\linewidth]{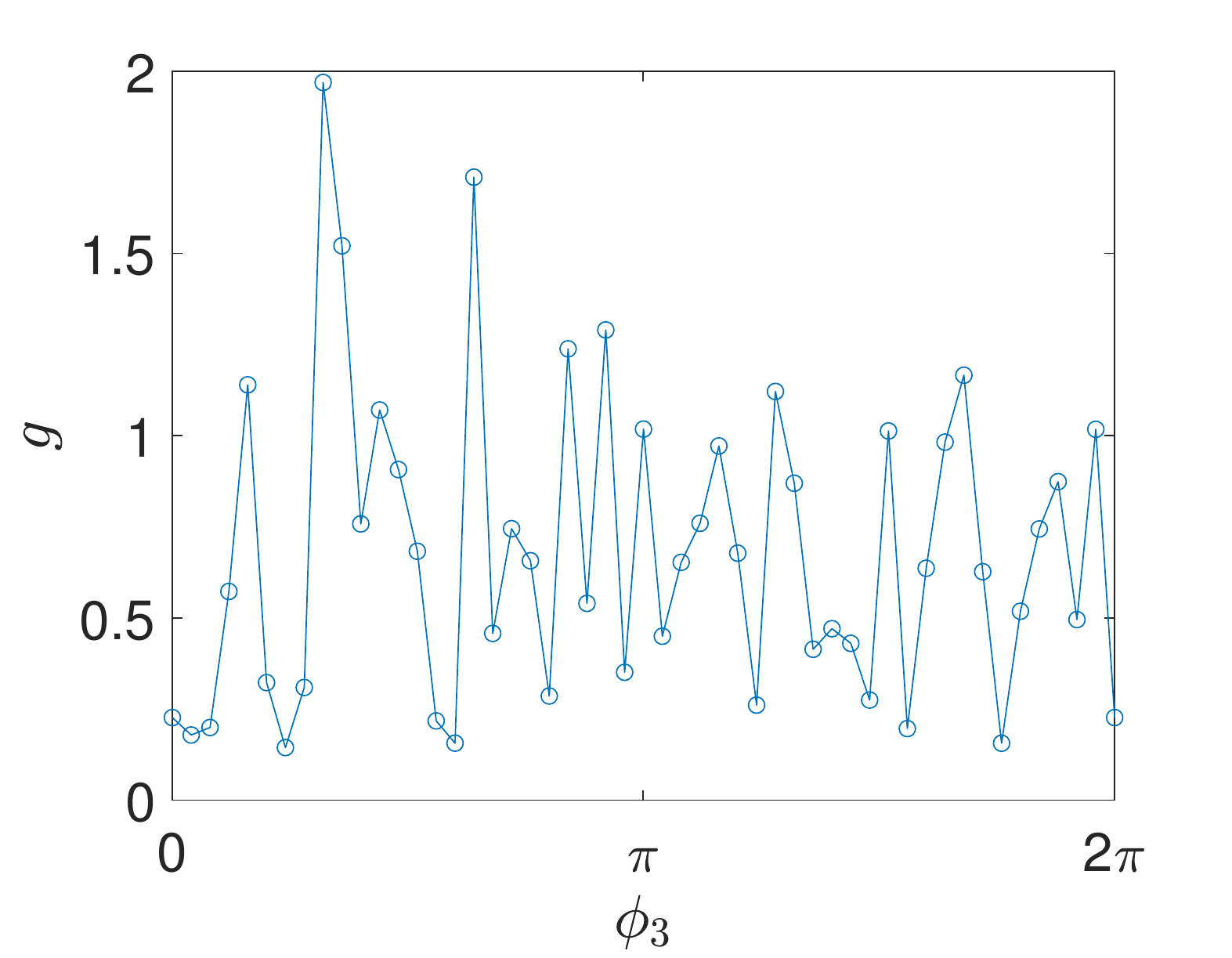}
				%{AAH_3D_couple_L20_V2_phi1_05_phi2_15_g.eps }
				%\caption{fig1}
			\end{minipage}%
		}%
		\subfigure{
			\begin{minipage}[t]{0.25\linewidth}
				\centering
				\includegraphics[width=1\linewidth]{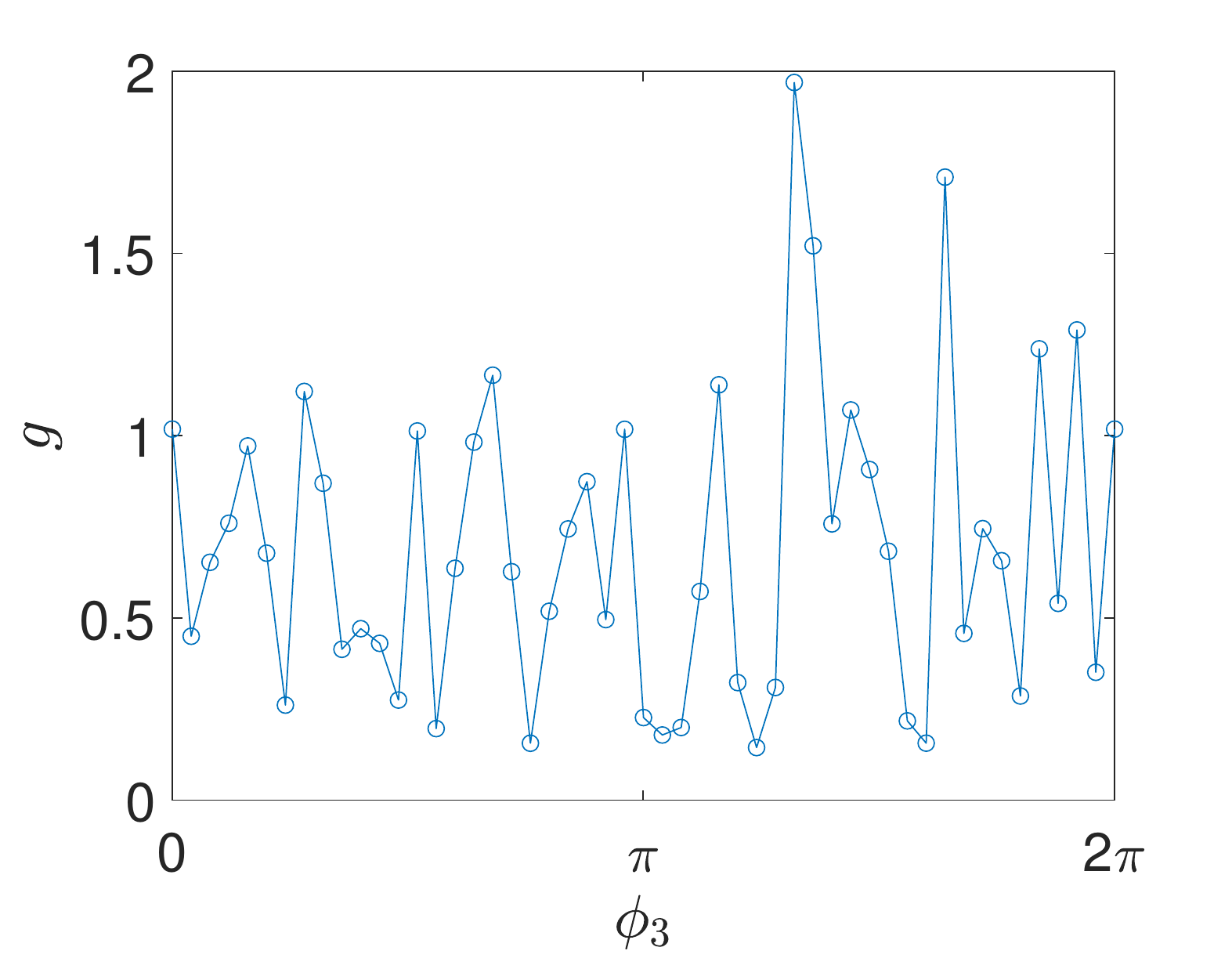}
				%{ AAH_3D_couple_L20_V2_phi1_15_phi2_05_g.eps }
				%\caption{fig1}
			\end{minipage}%
		}%	
		\subfigure{
			\begin{minipage}[t]{0.25\linewidth}
				\centering
				\includegraphics[width=1\linewidth]{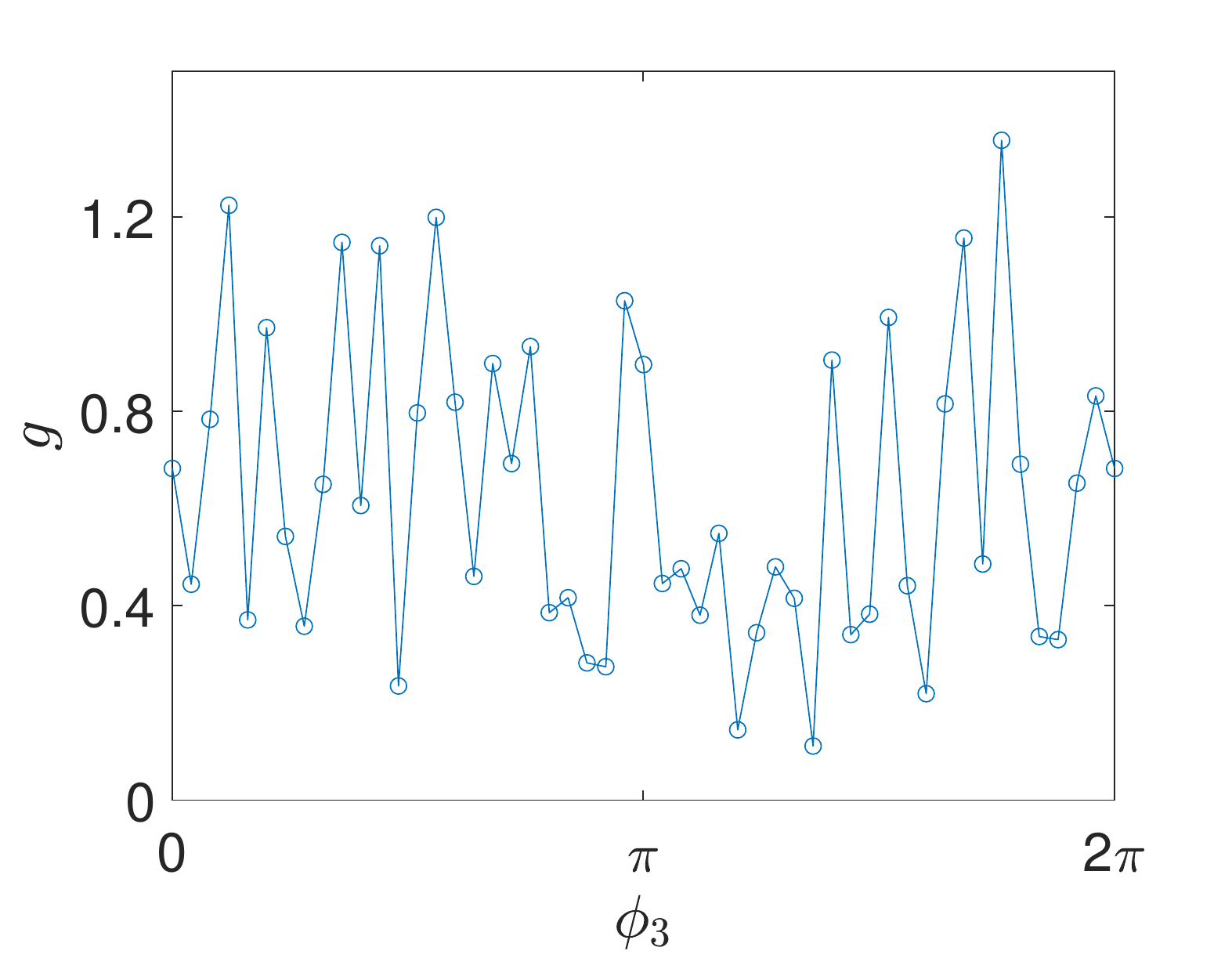}
				%{AAH_3D_couple_L20_V2_phi1_15_phi2_15_g.eps  }
				%\caption{fig1}
			\end{minipage}%
		}%
		
		\setcounter{subfigure}{0}
		\subfigure[ $\phi_1=0.5, \phi_2=0.5$]{
			\begin{minipage}[t]{0.25\linewidth}
				\centering
				\includegraphics[width=1\linewidth]{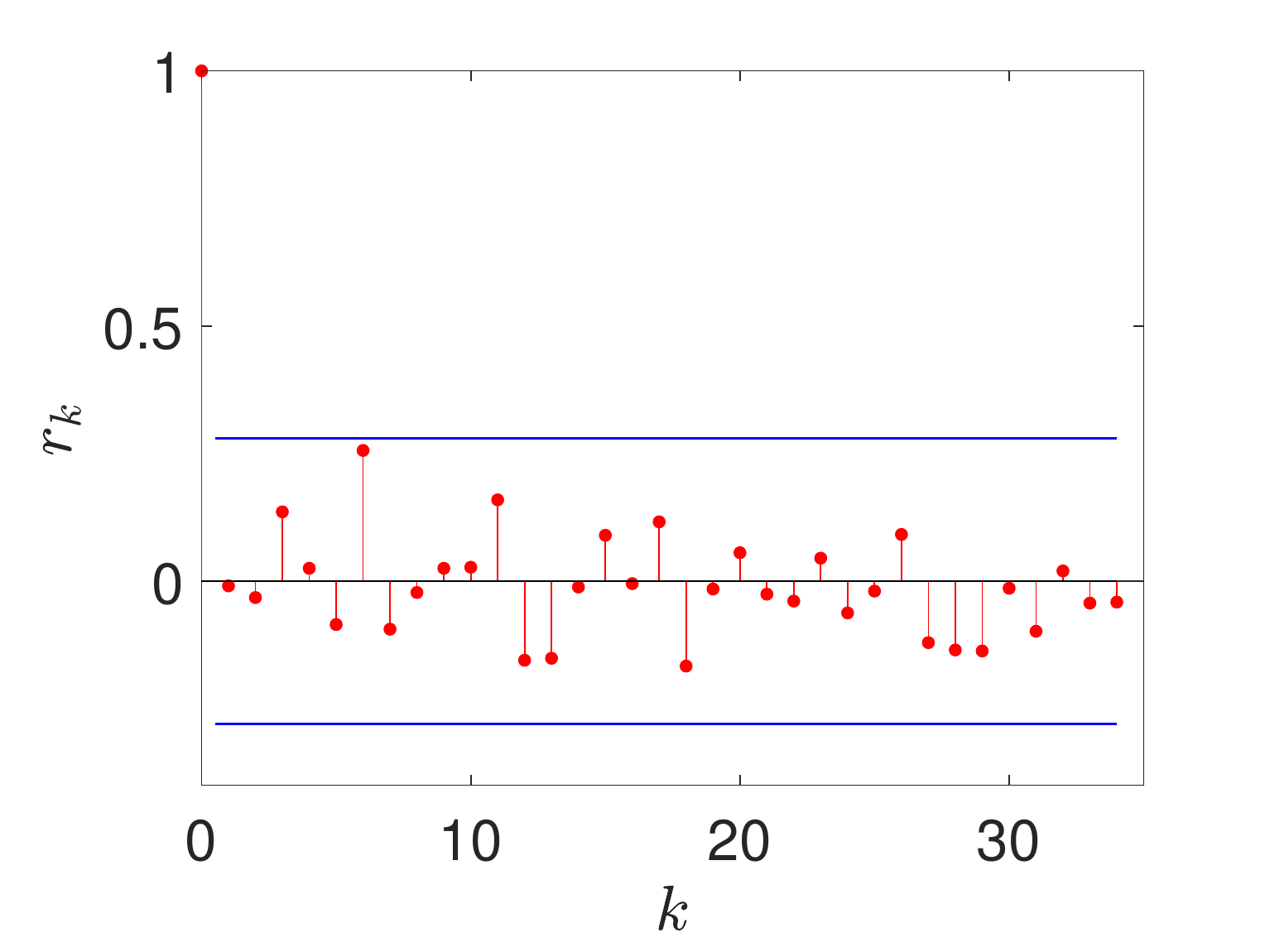}
				%{AAH_3D_couple_L20_V2_phi1_05_phi2_05.eps}
				%\caption{fig1}
			\end{minipage}%
		}%
		\subfigure[ $\phi_1=0.5, \phi_2=1.5$]{
			\begin{minipage}[t]{0.25\linewidth}
				\centering
				\includegraphics[width=1\linewidth]{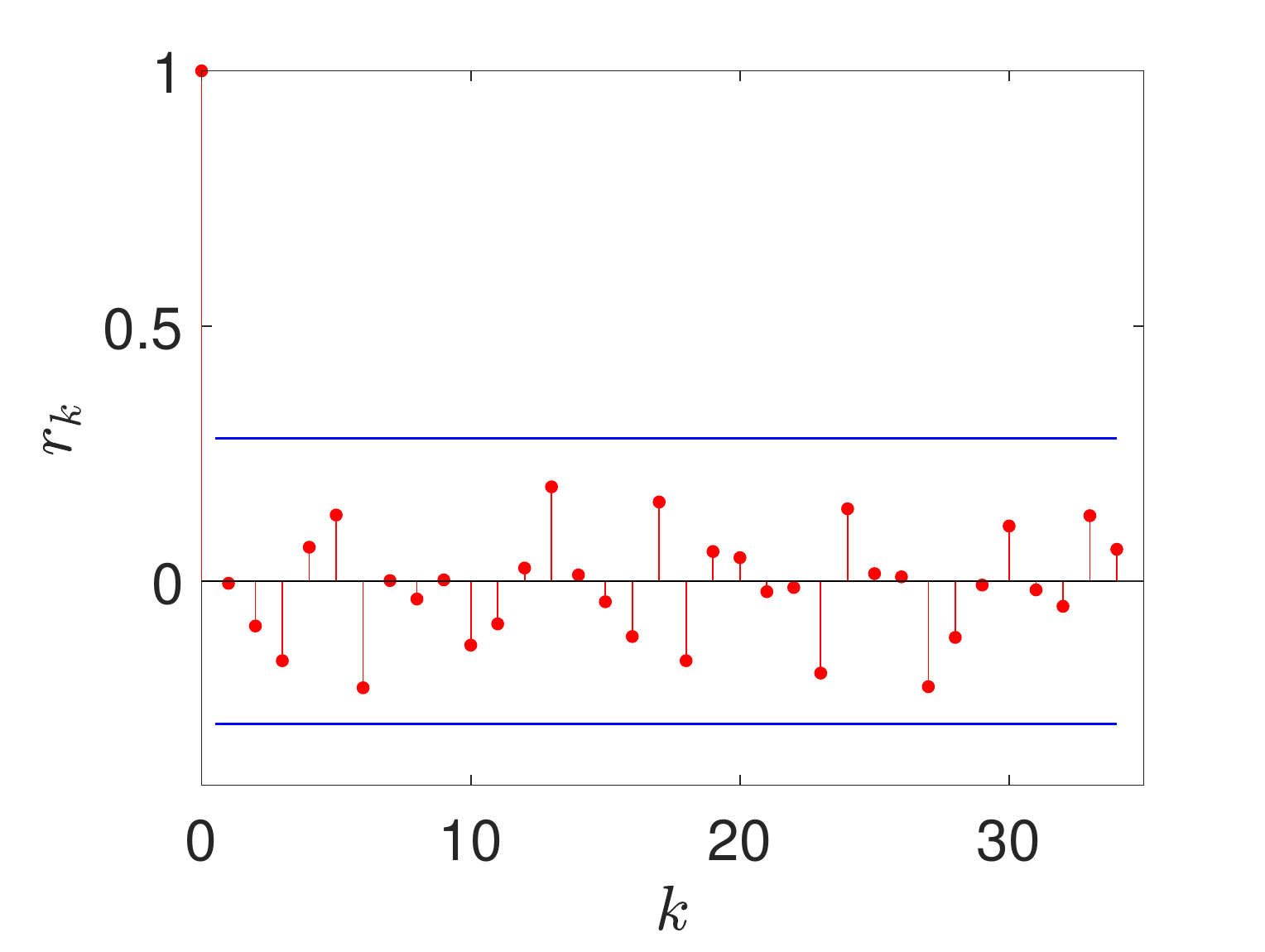}
				%{AAH_3D_couple_L20_V2_phi1_05_phi2_15.eps }
				%\caption{fig1}
			\end{minipage}%
		}%
		\subfigure[$\phi_1=1.5, \phi_2=0.5$]{
			\begin{minipage}[t]{0.25\linewidth}
				\centering
				\includegraphics[width=1\linewidth]{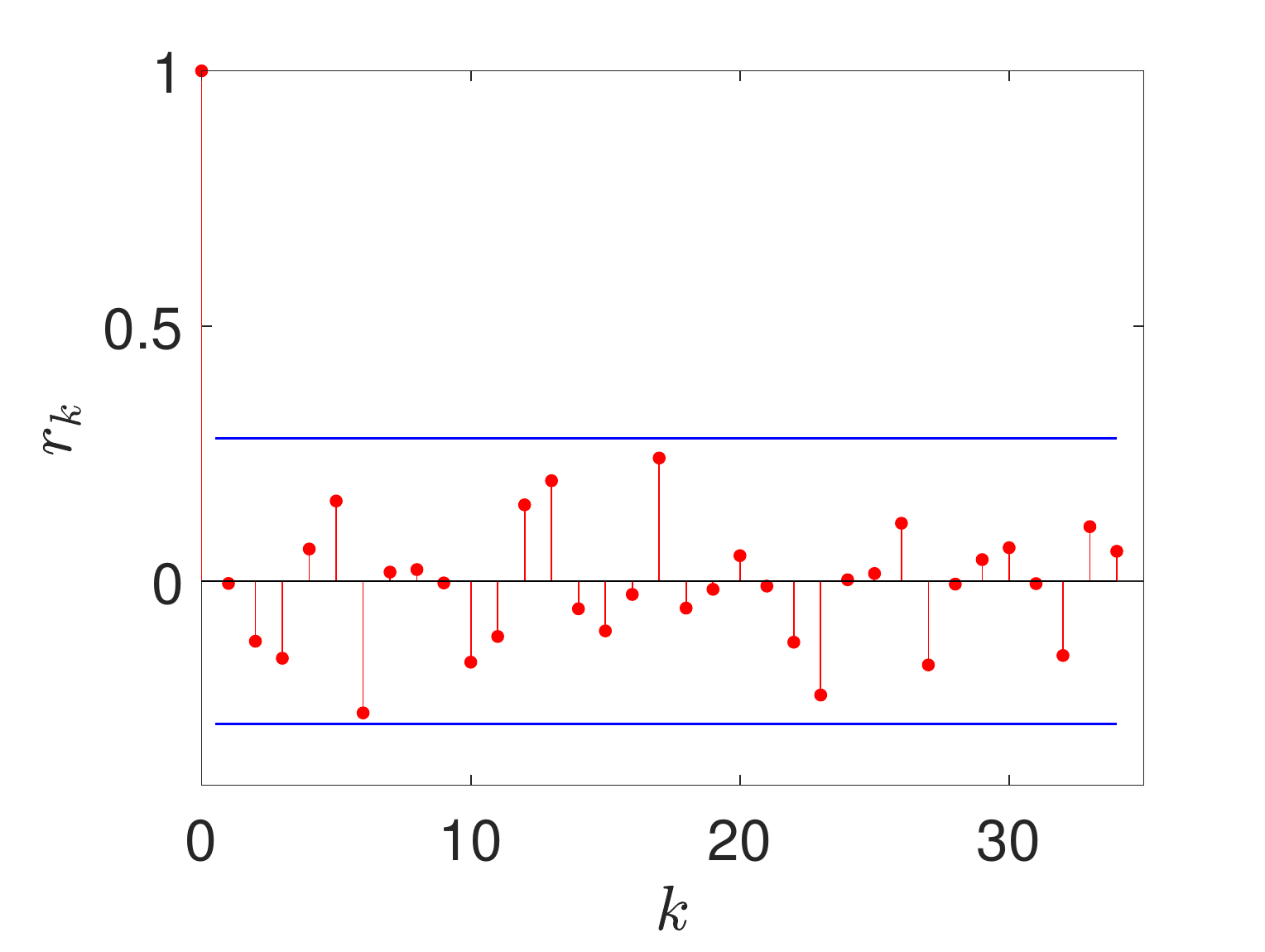}
				%{ AAH_3D_couple_L20_V2_phi1_15_phi2_05.eps }
				%\caption{fig1}
			\end{minipage}%
		}%	
		\subfigure[$\phi_1=1.5, \phi_2=1.5$]{
			\begin{minipage}[t]{0.25\linewidth}
				\centering
				\includegraphics[width=1\linewidth]{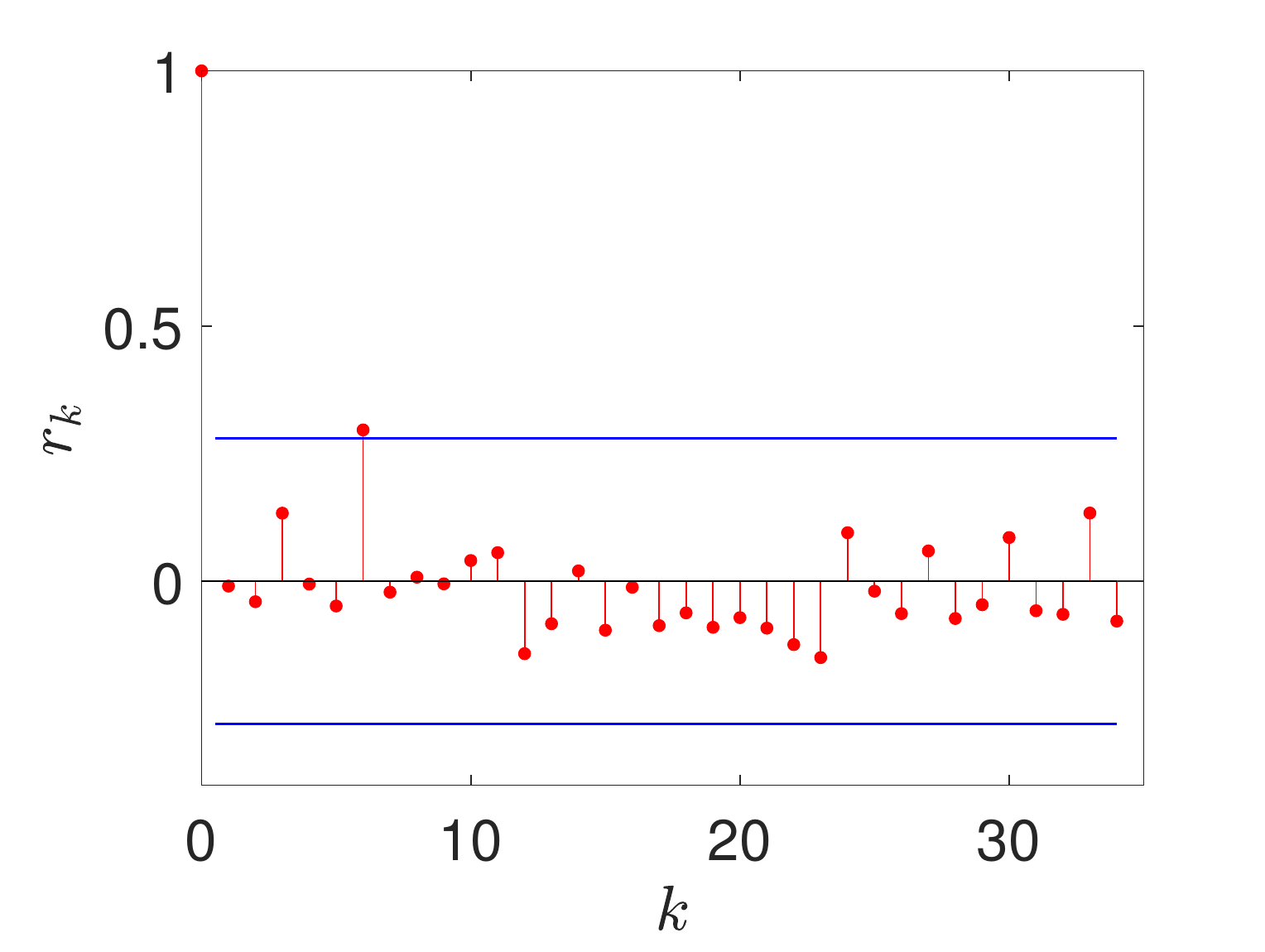}
				%{AAH_3D_couple_L20_V2_phi1_15_phi2_15.eps  }
				%\caption{fig1}
			\end{minipage}%
		}%
		
		\subfigure{
			\begin{minipage}[t]{0.25\linewidth}
				\centering
				\includegraphics[width=1\linewidth]{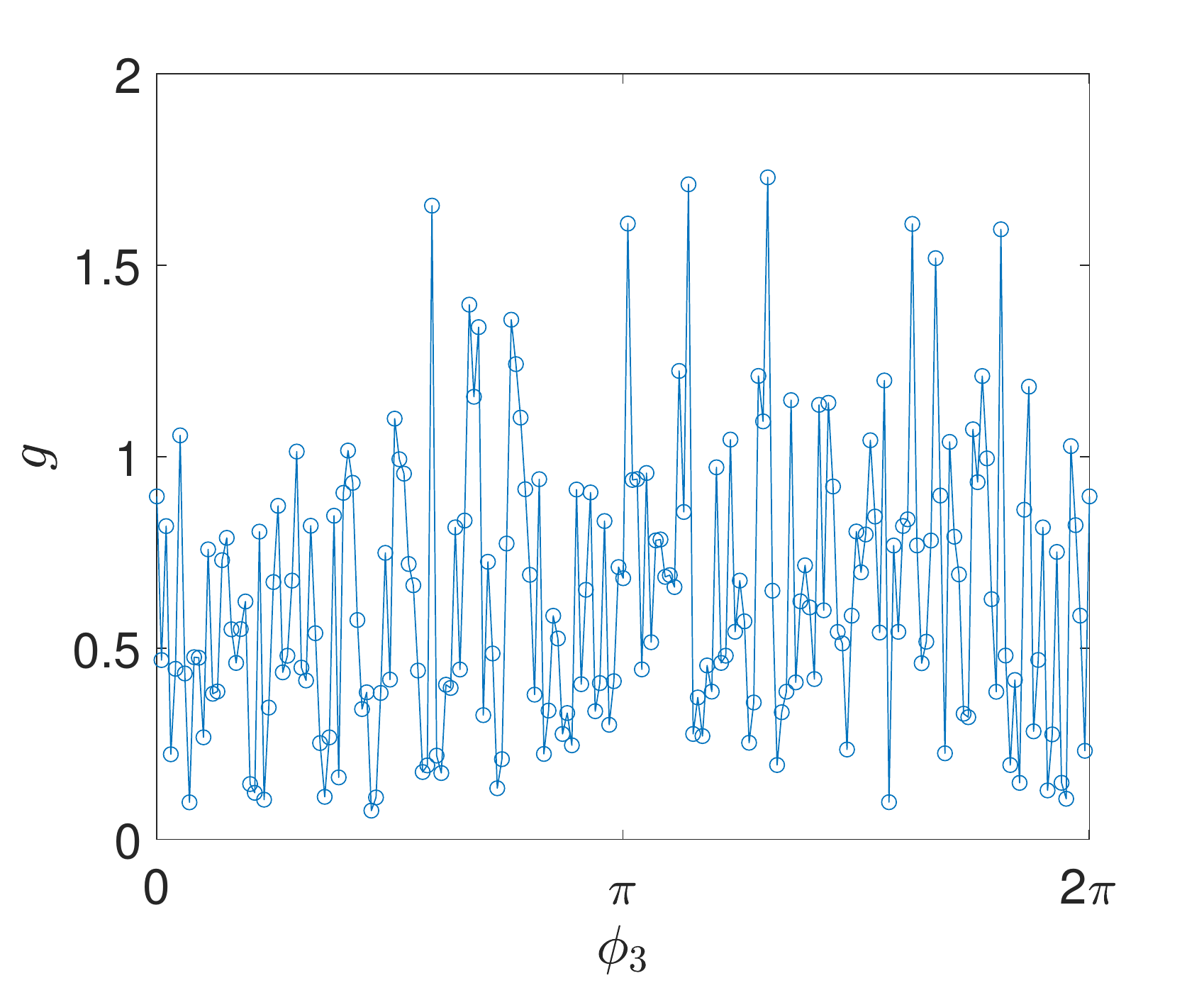}
				%{AAH_3D_couple_L20_V2_phi1_05_phi2_05_deltaPhi001pi_g.eps}
				%\caption{fig1}
			\end{minipage}%
		}%
		\subfigure{
			\begin{minipage}[t]{0.25\linewidth}
				\centering
				\includegraphics[width=1\linewidth]{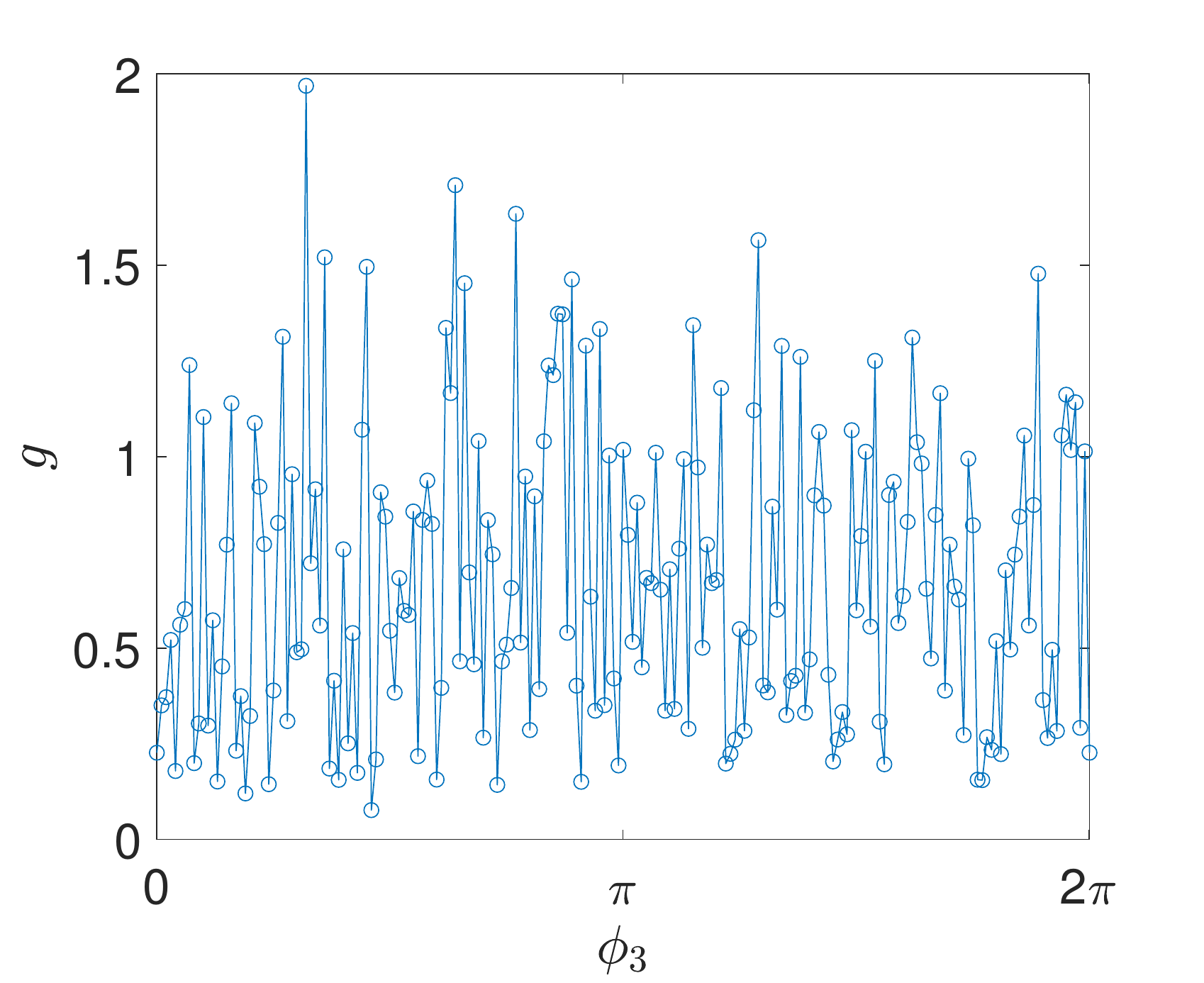}
				%{AAH_3D_couple_L20_V2_phi1_05_phi2_15_deltaPhi001pi_g.eps }
				%\caption{fig1}
			\end{minipage}%
		}%
		\subfigure{
			\begin{minipage}[t]{0.25\linewidth}
				\centering
				\includegraphics[width=1\linewidth]{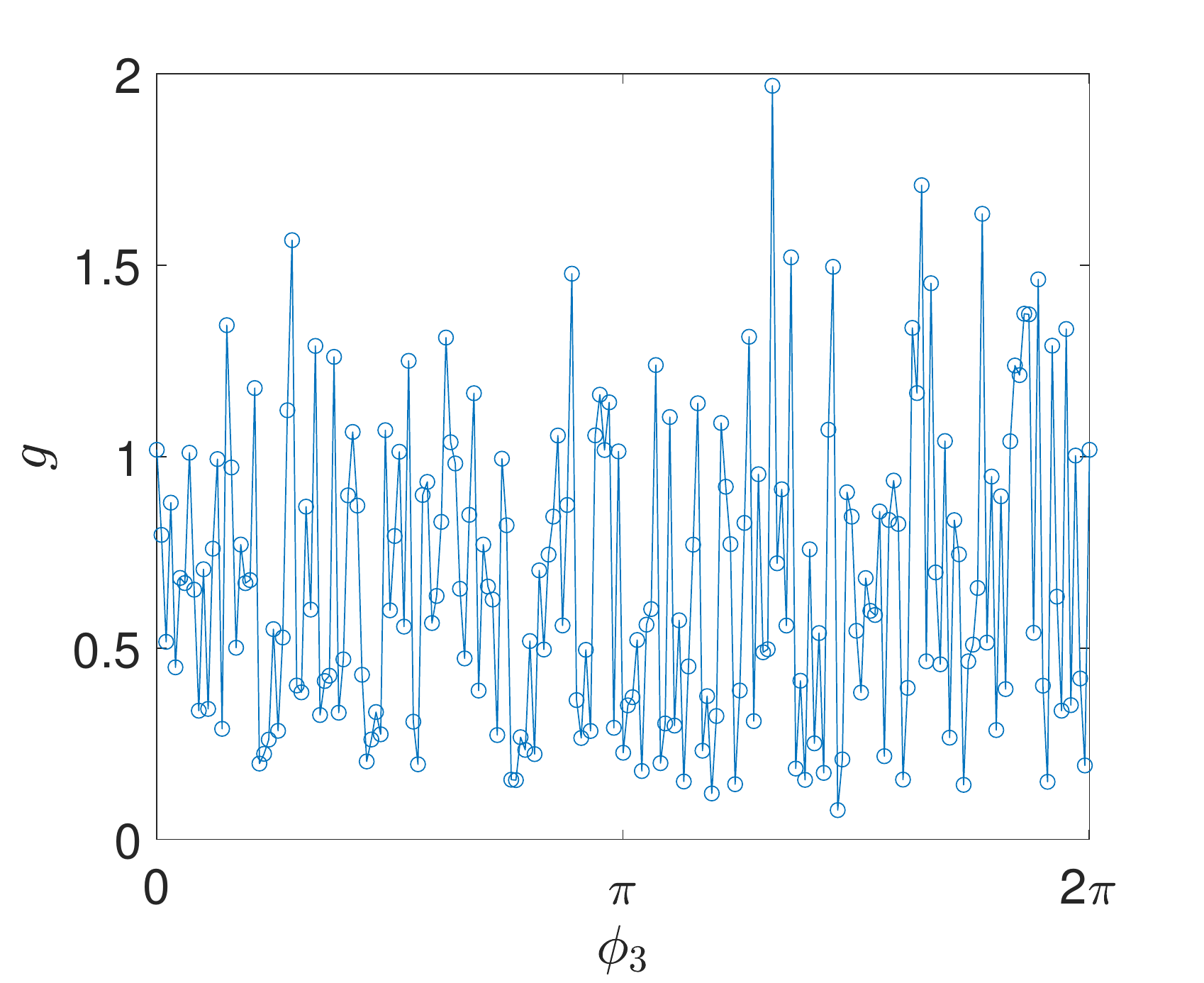}
				%{ AAH_3D_couple_L20_V2_phi1_15_phi2_05_deltaPhi001pi_g.eps }
				%\caption{fig1}
			\end{minipage}%
		}%	
		\subfigure{
			\begin{minipage}[t]{0.25\linewidth}
				\centering
				\includegraphics[width=1\linewidth]{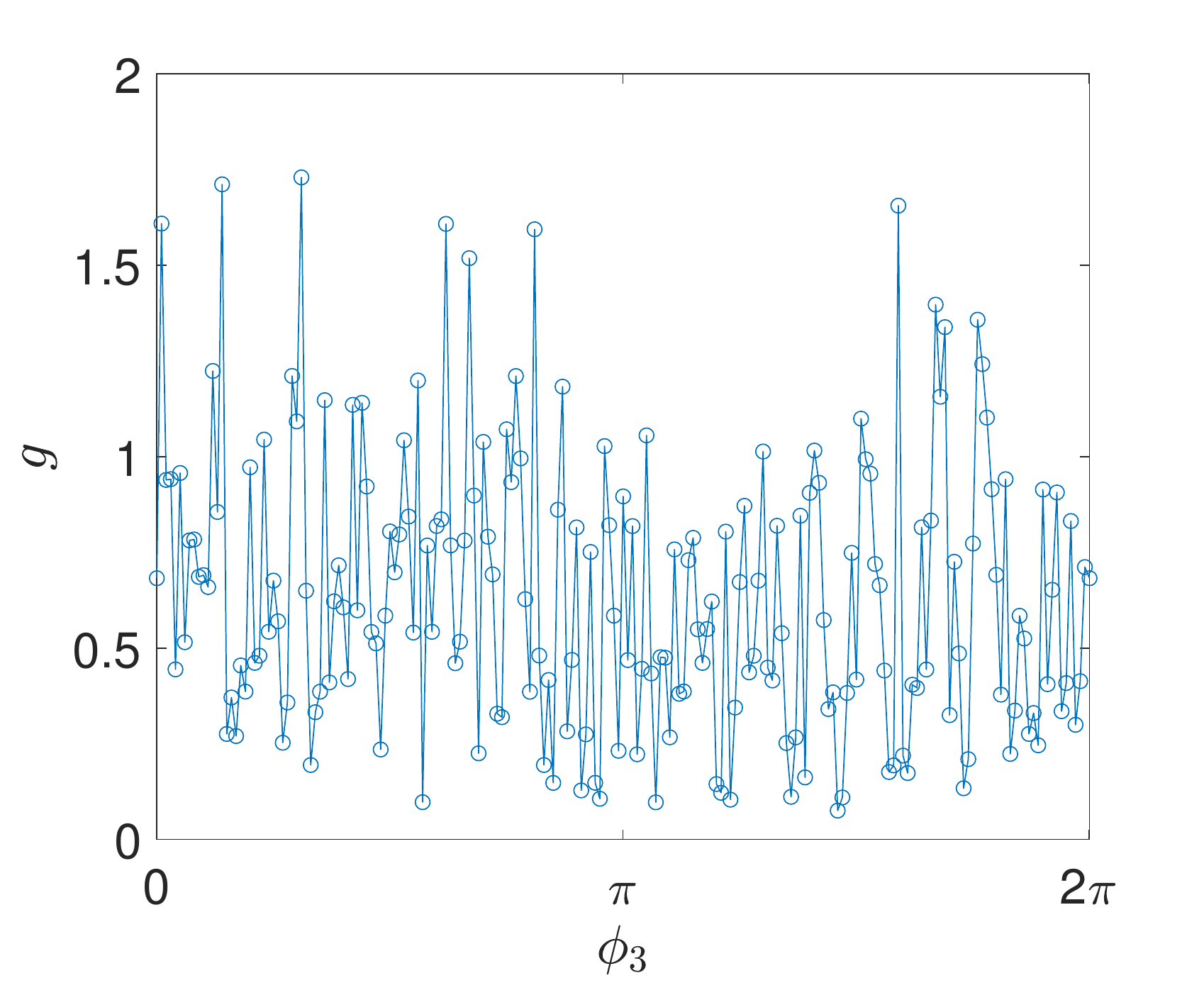}
				%{AAH_3D_couple_L20_V2_phi1_15_phi2_15_deltaPhi001pi_g.eps  }
				%\caption{fig1}
			\end{minipage}%
		}%
		\setcounter{subfigure}{4}
		\subfigure[ $\phi_1=0.5, \phi_2=0.5$]{
			\begin{minipage}[t]{0.25\linewidth}
				\centering
				\includegraphics[width=1\linewidth]{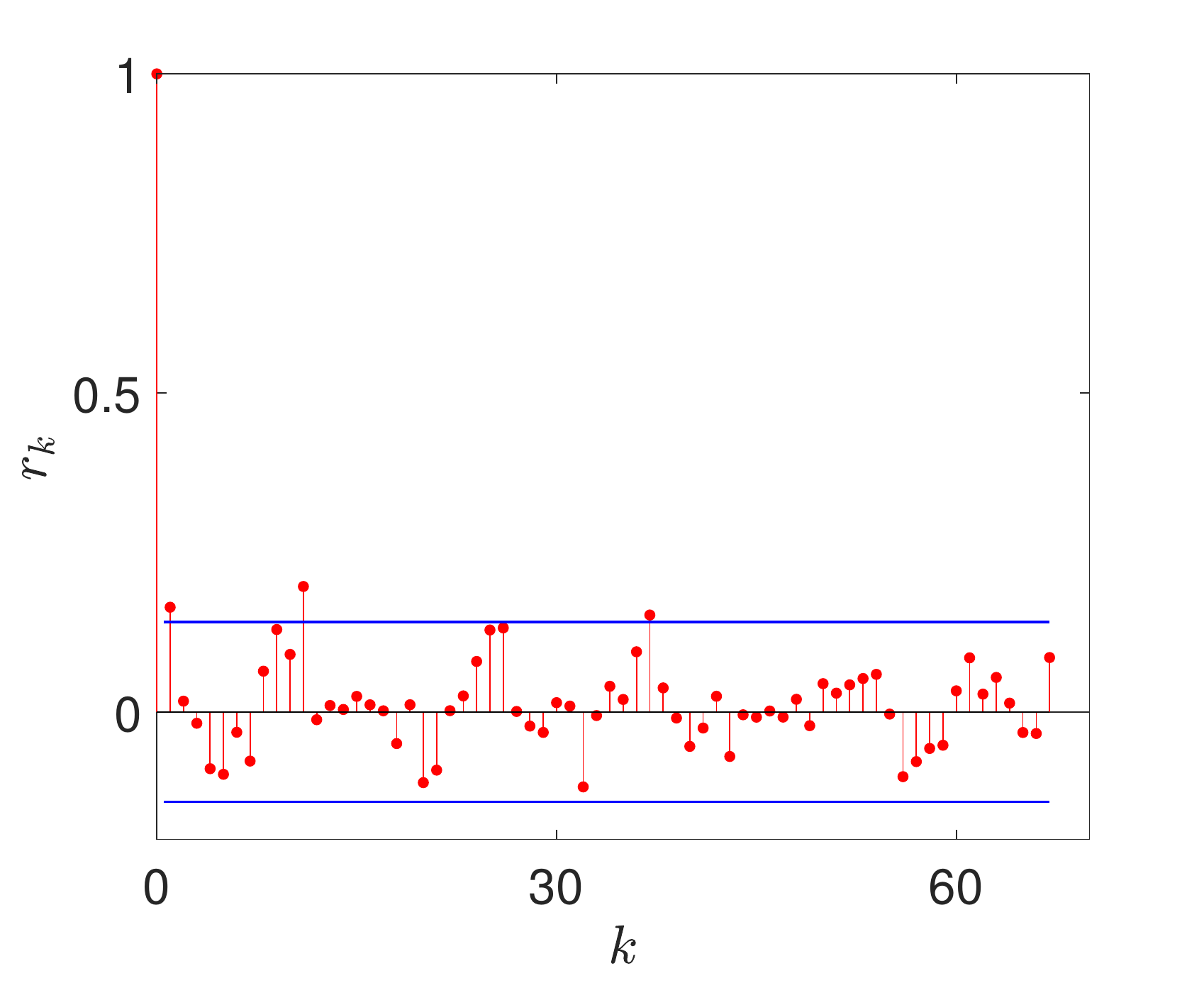}
				%{AAH_3D_couple_L20_V2_phi1_05_phi2_05_deltaPhi001pi.eps}
				%\caption{fig1}
			\end{minipage}%
		}%
		\subfigure[ $\phi_1=0.5, \phi_2=1.5$]{
			\begin{minipage}[t]{0.25\linewidth}
				\centering
				\includegraphics[width=1\linewidth]{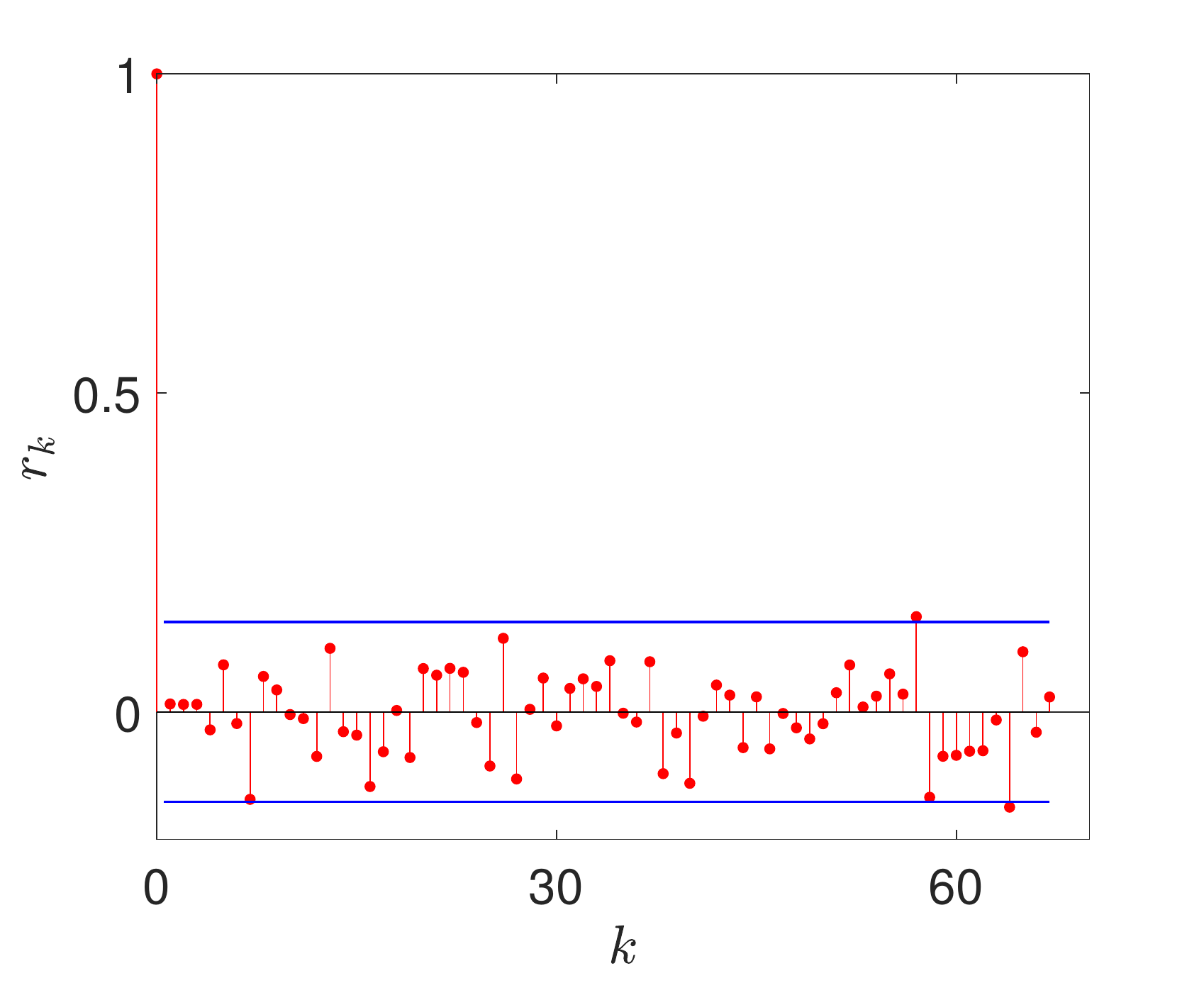}
				%{AAH_3D_couple_L20_V2_phi1_05_phi2_15_deltaPhi001pi.eps }
				%\caption{fig1}
			\end{minipage}%
		}%
		\subfigure[$\phi_1=1.5, \phi_2=0.5$]{
			\begin{minipage}[t]{0.25\linewidth}
				\centering
				\includegraphics[width=1\linewidth]{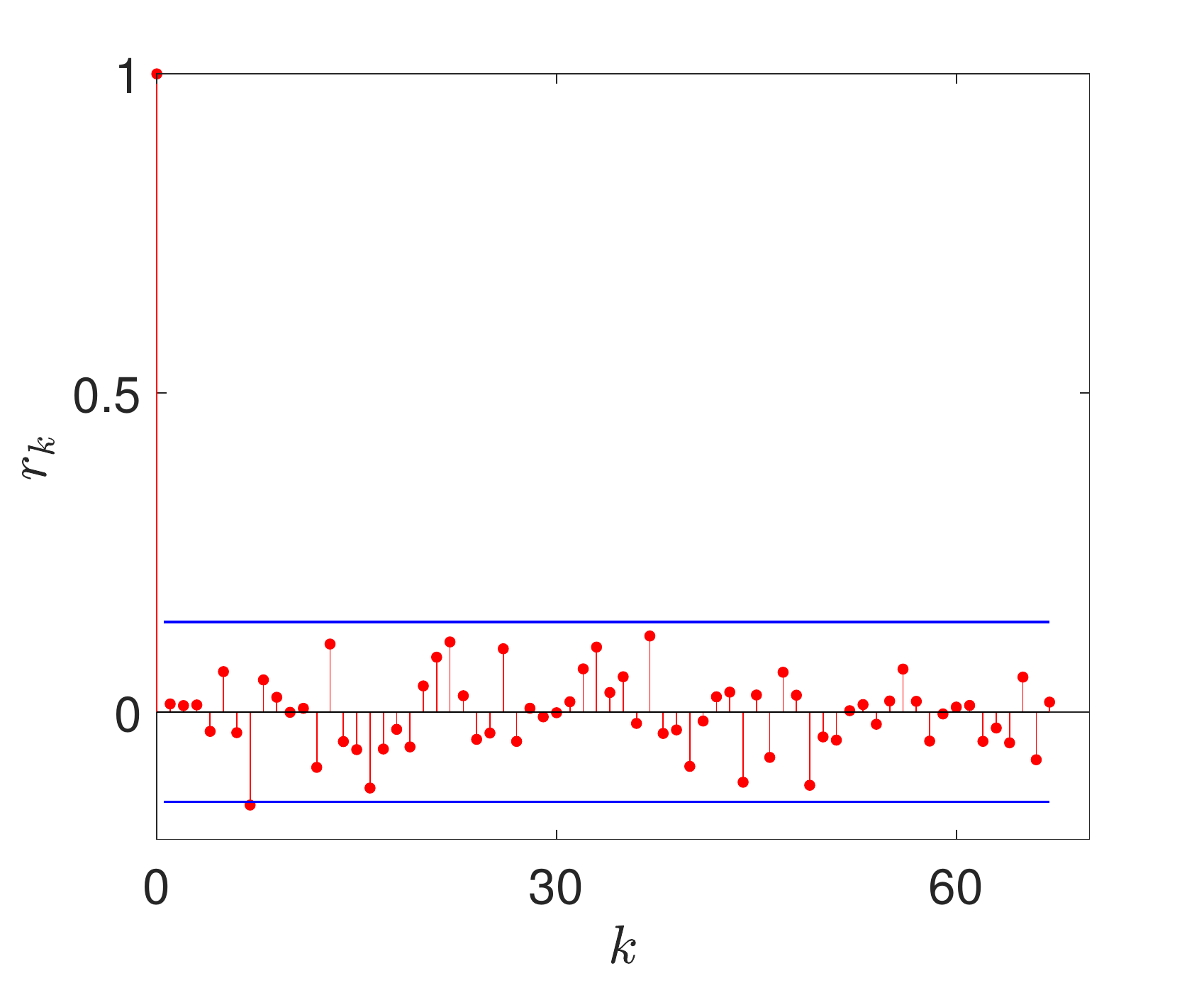}
				%{ AAH_3D_couple_L20_V2_phi1_15_phi2_05_deltaPhi001pi.eps }
				%\caption{fig1}
			\end{minipage}%
		}%	
		\subfigure[$\phi_1=1.5, \phi_2=1.5$]{
			\begin{minipage}[t]{0.25\linewidth}
				\centering
				\includegraphics[width=1\linewidth]{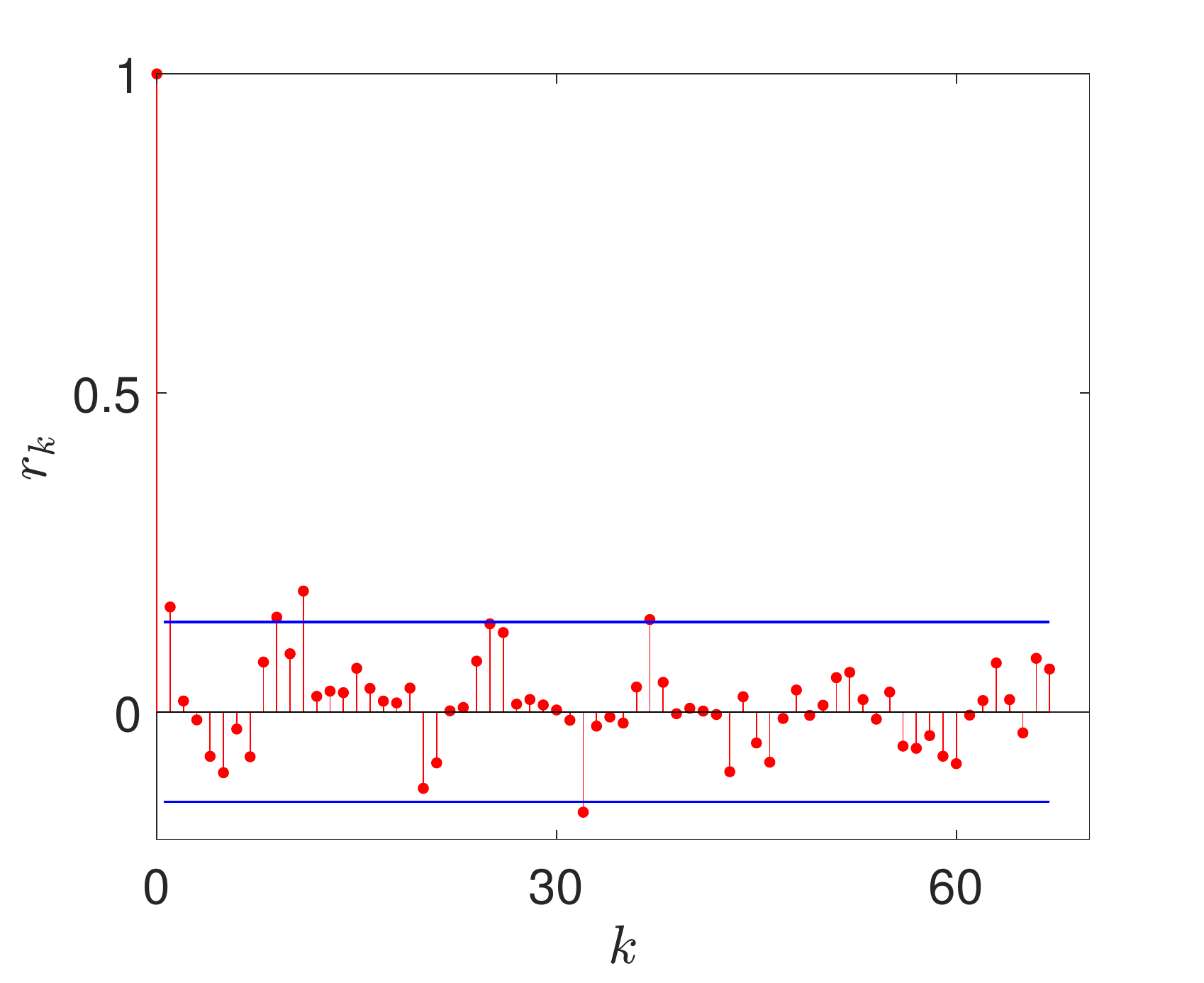}
				%{AAH_3D_couple_L20_V2_phi1_15_phi2_15_deltaPhi001pi.eps  }
				%\caption{fig1}
			\end{minipage}%
		}%
		\caption{The upper panel of (a)-(h): the conductance as a function of $\phi_3$ with $\phi_3$ changed from zero to $2\pi$ with the interval $\delta_{\phi}=0.04\pi$ for (a)-(d), and
			$\delta_{\phi}=0.01\pi$ for (e)-(h). $\phi_1$ and $\phi_2$ are fixed.
			The conductance is calculated along the $z$ direction for the Anderson model with quasiperiodic potential strength $V=2$ in the cubic system $L_x=L_y=L_z=20$.
			The lower panel of (a)-(h): the autocorrelation $r_k$ as a function of lag $k$ for the conductance in the upper panel.
			The red-filled circles are the autocorrelation and 
			the two blue lines are the two standard error limits indicating the $95\%$ confidence interval under the assumption that the series is completely random.}
		\label{fig_autocorr_g}
	\end{figure*}
	
	\section{Autocorrelation}\label{Appendix_ACF}
	To test whether a series of data is correlated or not, let us review how the autocorrelation
	is defined and calculated.
	Suppose there is a series of data sorted by time $t$, i.e., $\{y_t\}$ with $t=1,2,\cdots,T$.
	The autocorrelation measures the correlation between $y_t$ and $y_{t + k}$
	with $k = 0,1,2,...,K$.
	According to Ref. \onlinecite{Hamilton94}, the autocorrelation for lag $k$ is defined as
	\begin{align}
		r_k\equiv\frac{c_k}{c_0},
	\end{align}
	where
	\begin{align}
		c_k\equiv\frac{1}{T}\sum_{t=1}^{T-k}(y_t-\langle y\rangle) (y_{t+k}-\langle y\rangle).
	\end{align}
	$\langle \cdot\rangle$ stands for the mean value in statistics and $c_0$ is the sample variance of the time series.
	
	Suppose that $q$ is the lag beyond which the theoretical autocorrelation is effectively zero. 
	Then, according to Ref. \onlinecite{Hamilton94}, the estimated standard error of the autocorrelation at lag $k>q$ is
	\begin{align}
		\sigma_{r_k}=\sqrt{\frac{1}{T}(1+\sum_{j=1}^{q}r_j^2)}.
	\end{align}
	If the series is completely random, then the standard error reduces to $1/\sqrt{T}$.
	Two standard error limits determine the $95\%$ confidence interval under the assumption that the series is completely random.
	If all of the autocorrelation for lag $k>0$ is within the $95\%$ confidence interval,
	then we can accept the assumption, otherwise, we should reject the assumption.
	
	In this paper, the time is replaced with the sample index,
	and the time $T$ will be replaced by the number of samples $N$.

	\section{Autocorrelation of conductance}\label{Appendix_auto_g}
	The conductance has been averaged by $N$ samples with random phase $\{\phi_i\}$;
	\begin{align}
		\langle g\rangle\equiv \frac{1}{N}\sum_i^{N}g_i.
	\end{align}
	Then the standard error of $\langle g\rangle$ has been estimated as follows,
	\begin{align}
		\sigma_{\langle g\rangle}=\frac{\sigma_{g}}{\sqrt{N}}, \label{sigma_g}
	\end{align}   
	where $\sigma_g$ is the standard deviation for the single sample,
	\begin{align}
		\sigma_{g}=\sqrt{\frac{\sum_{i=1}^N (g_i-\langle g\rangle)^2}{N-1}}.
	\end{align}
	Eq. (\ref{sigma_g}) is correct when each $g_i$ is independent and obeys the same probability distribution.
	However, in the quasiperiodic systems of this paper, 
	the conductance will be almost the same when the two sets of $\{\phi_i\}$ are close to each other.
	Then it would produce the correlation between the conductance when $\{\phi_i\}$'s are close to each other.
	We use $\{\delta_{\phi_{i}}\}$ to characterize the distance between $\{\phi_i\}$ and
	its nearest neighbor. $\{\delta_{\phi_{i}}\}$ should be large enough to guarantee 
	there is no correlation between samples. The allowed minimum
	% \textcolor{red}{(I changed maximum to minimum)} 
	distance is labeled as $\delta_{\phi}$.
	In the actual calculations,
	we simulate $N$ samples in the space of $\phi_1-\phi_2-\phi_3$ with $\phi_i\in [0,2\pi), i=1,2,3$, 
	then the allowed simulated samples  satisfy
	\begin{align}
		N<(2\pi/\delta_{\phi})^3. \label{N_restrict}
	\end{align}

	In order to estimate $\delta_{\phi}$, we check the correlation of conductance with $\{\phi_{i}\}$
	in a linear line in the space of $\phi_1-\phi_2-\phi_3$.
	For example, 
	we fix $\phi_1$ and $\phi_2$, and change $\phi_3$. 
	Then the conductance $g$ is a continuous function of $\phi_3$, which varies smoothly
	within $\delta_{\phi}$. 
	We use the autocorrelation introduced in Appendix \ref{Appendix_ACF} to test whether there is any correlation.
	From Fig. \ref{fig_autocorr_g} (a)-(d), it is safe to say there is no correlation when $\delta_{\phi}=0.04\pi$.
	However, the correlation is obvious when $\delta_{\phi}=0.01\pi$ 
	[Fig. \ref{fig_autocorr_g} (e)-(h)].
	Then the allowed number of samples should satisfy $N<1.25\times10^5$. 
	We, therefore, chose $N=50000$ for the AM.
	We also checked the $\delta_{\phi}$ for the PPM and Ando model
	and found that the number of samples set in this paper satisfies Eq. (\ref{N_restrict}).

	\begin{figure}[bt]
		\centering
		\subfigure[$L=12, L_z=10^4$ ]{
			\begin{minipage}[t]{0.5\linewidth}
				\centering
				\includegraphics[width=1\linewidth]{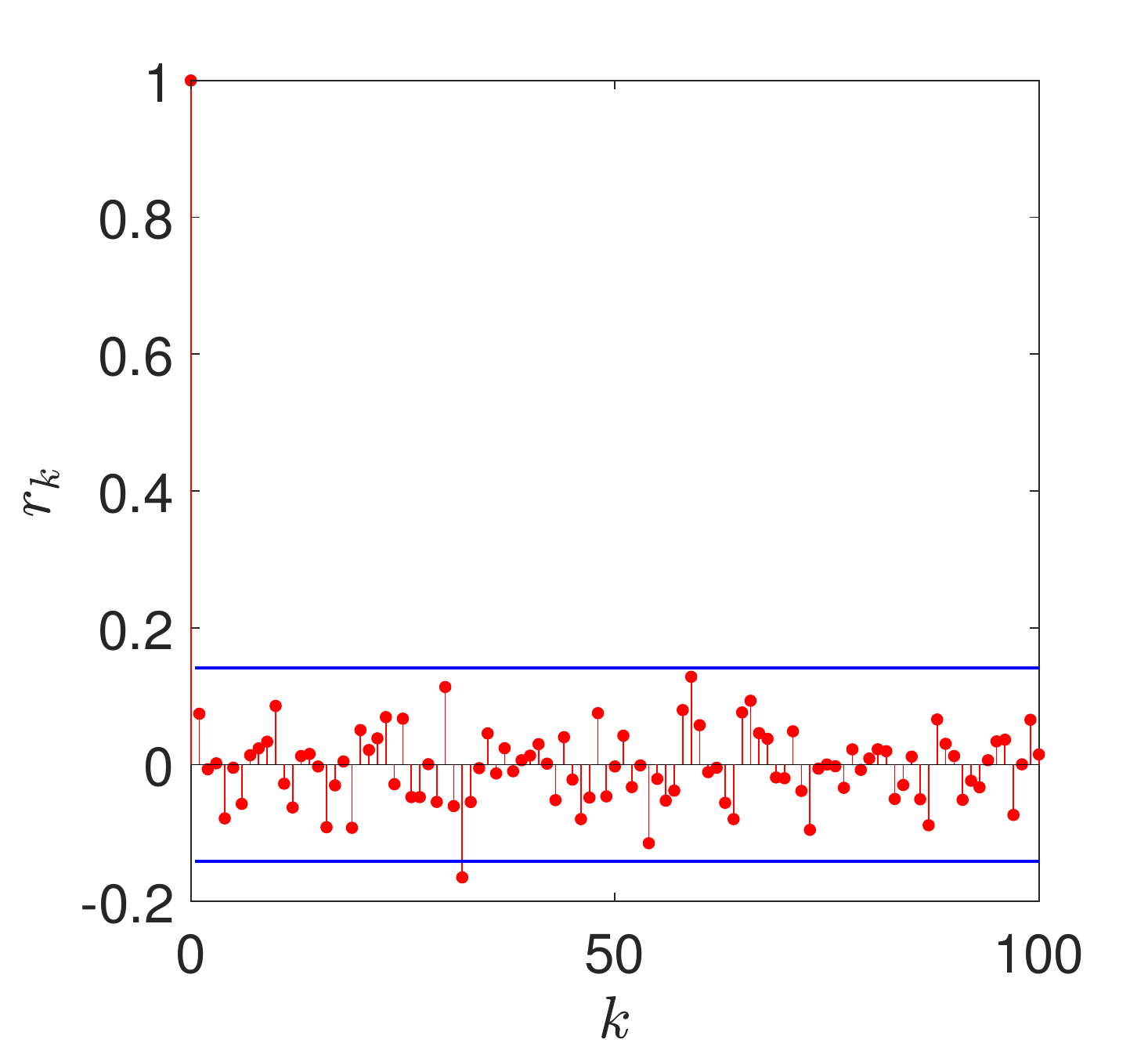}
				%{ACF_AAH_3D_couple_Single_LE_E=0_V=2_L=12_Lz=10000.eps}
				%\caption{fig1}
			\end{minipage}%
		}%
		\subfigure[$L=12, L_z=10^5$ ]{
			\begin{minipage}[t]{0.5\linewidth}
				\centering
				\includegraphics[width=1\linewidth]{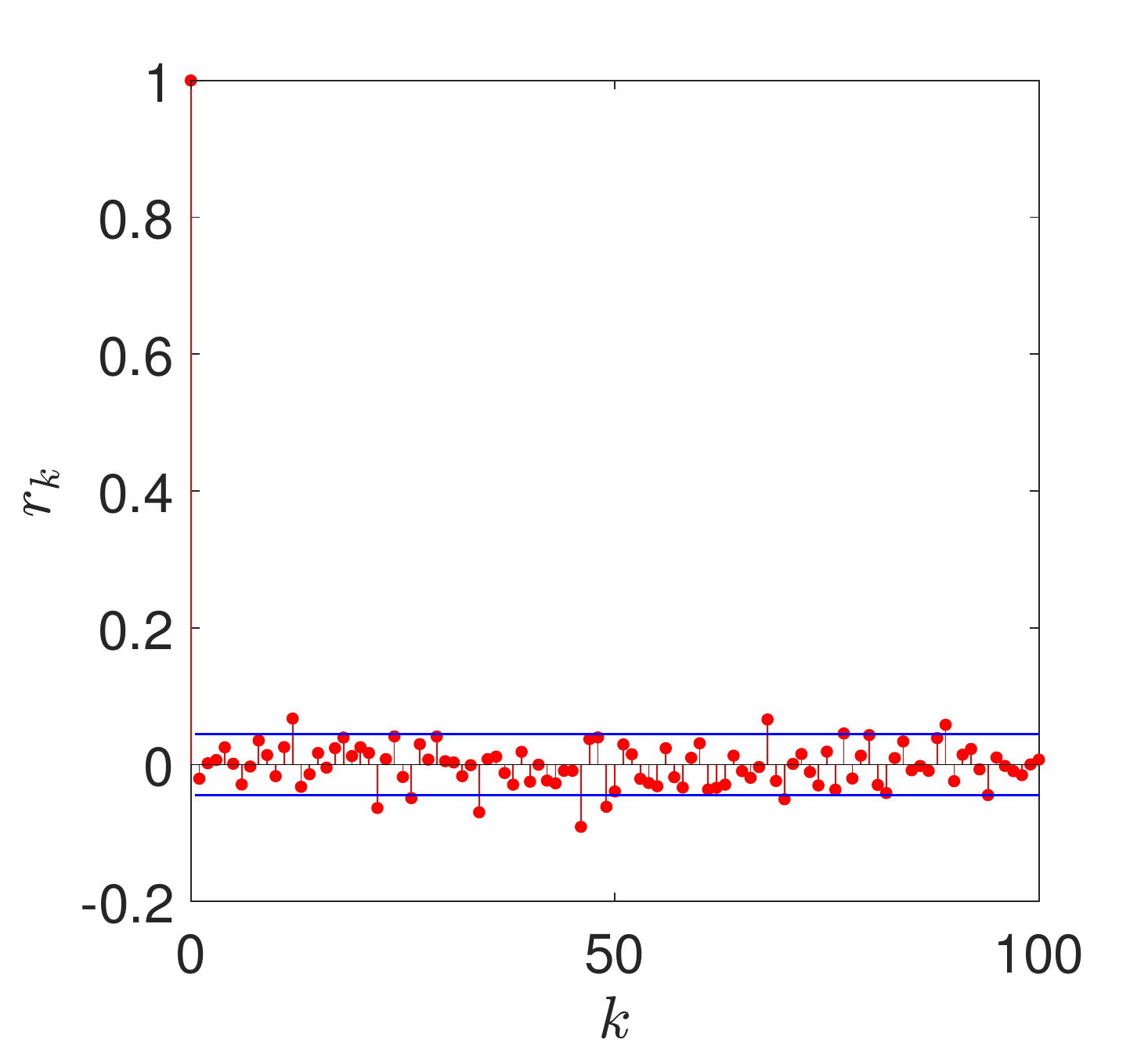}
				%{ACF_AAH_3D_couple_Single_LE_E=0_V=2_L=12_Lz=100000.eps}
				%\caption{fig1}
			\end{minipage}%
		}%
		
		\subfigure[$L=12, L_z=10^6$ ]{
			\begin{minipage}[t]{0.5\linewidth}
				\centering
				\includegraphics[width=1\linewidth]{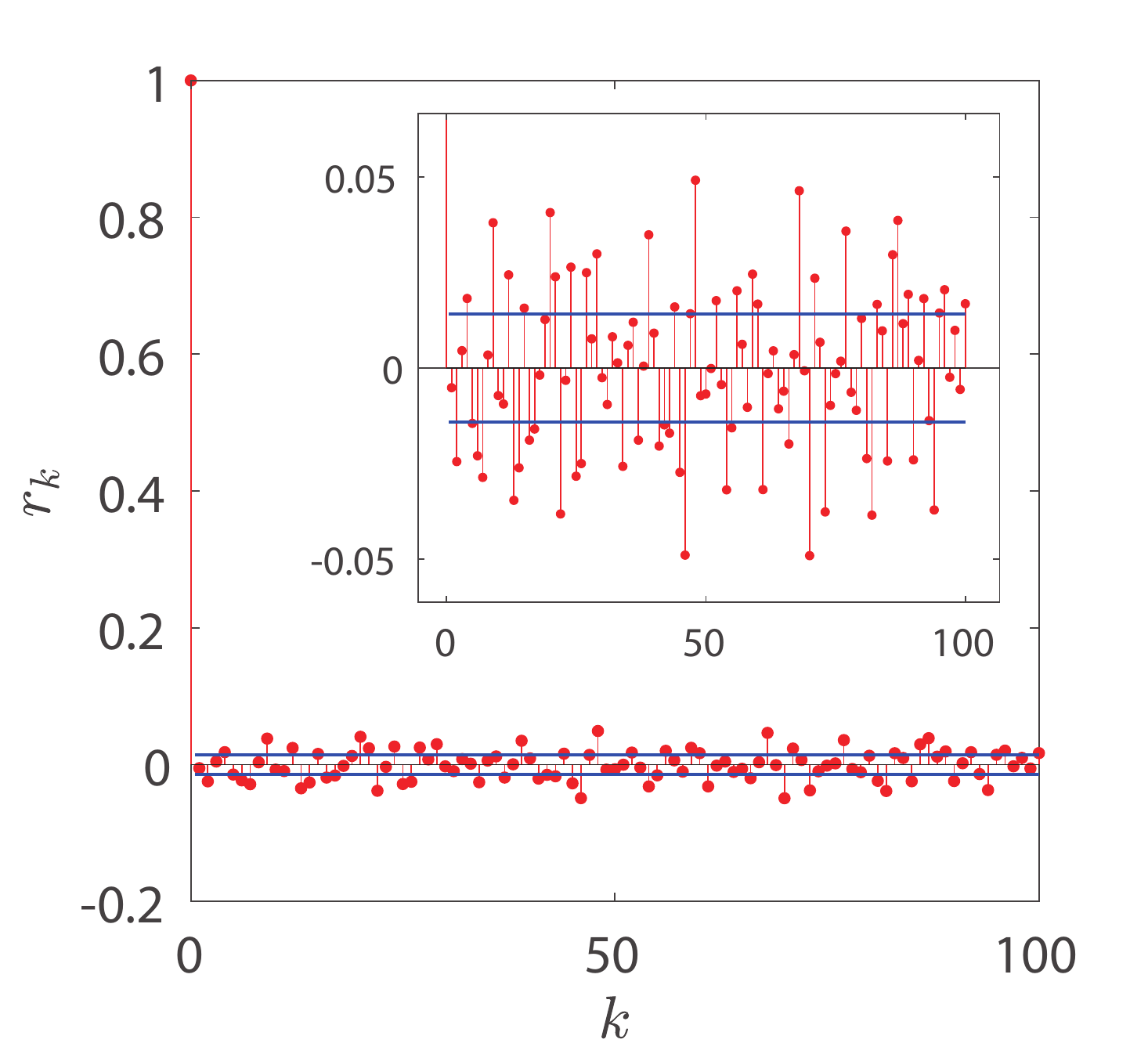}
				%{ACF_AAH_3D_couple_Single_LE_E=0_V=2_L=12_Lz=1000000.eps}
				%\caption{fig1}
			\end{minipage}%
		}%
		\subfigure[$L=16, L_z=10^6$ ]{
			\begin{minipage}[t]{0.5\linewidth}
				\centering
				\includegraphics[width=1\linewidth]{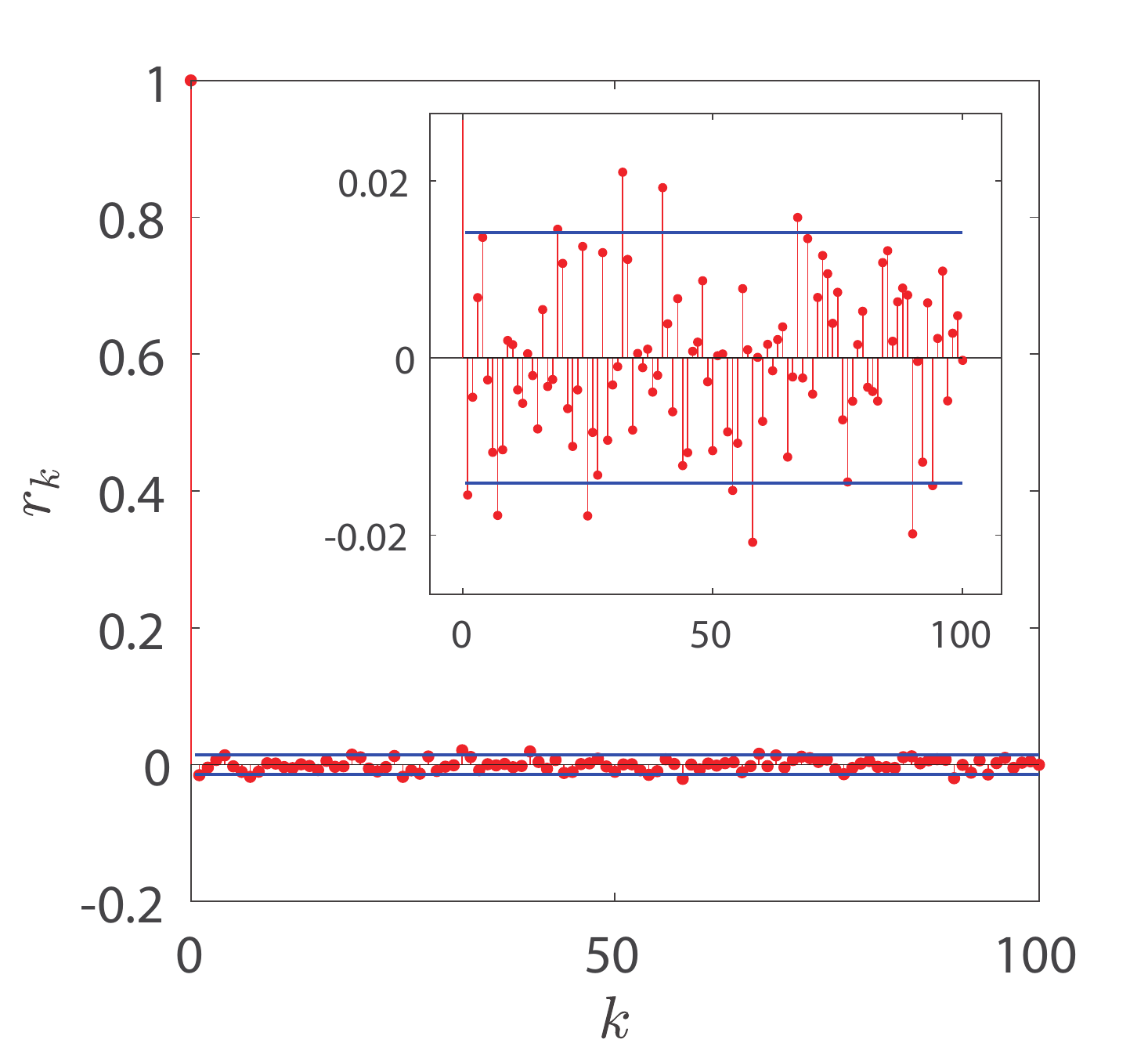}
				%{ACF_AAH_3D_couple_Single_LE_E=0_V=2_L=16_Lz=1000000.eps}
				%\caption{fig1}
			\end{minipage}%
		}%
		\caption{The autocorrelation $r_k$ as a function of lag $k$ for
			the consecutive series of $\{\gamma_i\}$ in the quasi-1D Anderson model with
			cross-section size $L$ and transmission length $L_z$.
			The red-filled circles are the autocorrelations and the two blue lines are the two standard error limits 
			determines the $95\%$ confidence interval under the assumption that the series is completely random.
			Insets in (c) and (d): the enlarged pictures to show the detail except $r_0=1$.
		}
		\label{autocorr_LE}
	\end{figure}
	
	\section{Autocorrelation of Lyapunov exponent}\label{Appendix_auto_lambda}
	
	The localization length $\lambda\equiv1/\gamma$ with $\gamma$ the Lyapunov exponent
	is also a good quantity to characterize the Anderson transition. 
	However, we find there is also a problem with the error estimation because of the long-range correlation of QP.  
	
	Let us first review the method to estimate the error of the Lyapunov exponent \cite{Slevin14}.
	We focus on the wavefunction decay in a quasi-1D system with $L_x=L_y=L\ll L_z$.
	The localization length is the slowest decay of the amplitude of wavefunction 
	and is related to the smallest positive Lyapunov exponent $\gamma$ by $\lambda\equiv1/\gamma$.
	In practice, we do the $QR$ decomposition every $q$ layer of multiplications of the transfer matrix.  
		We treat $p$ ($=n\times q$ with $n$ integer) layers as a statistics independent sample, 
		and the $i$th sample gives the estimated smallest positive Lyapunov exponent $\gamma_{i}$.
	Then the smallest positive Lyapunov exponent is estimated as
	\begin{align}
		\gamma=\frac{1}{N}\sum_i^N\gamma_i
	\end{align}
	with $N=L_z/p$. 
	Then their sample standard deviation is estimated as,
	\begin{align}
		\sigma=\sqrt{\frac{\sum_{i=1}^N (\gamma_i-\gamma)^2}{N-1}}
	\end{align}
	and the standard deviation of $\gamma$ is estimated as,
	\begin{align}
		\sigma_{ \gamma }=\frac{\sigma}{\sqrt{N}}. \label{sigma_gamma}
	\end{align}
	
	Similar to the conductance, $\{\gamma_{i}\}$ should be independent and without correlations
		in order to estimate $\sigma_{ \gamma }$ correctly according to Eq. (\ref{sigma_gamma}).
	However, we find there are correlations for $\{\gamma_{i}\}$ in our quasiperiodic systems.
	We use the autocorrelation introduced in Appendix \ref{Appendix_ACF} to test the correlation of a series of $\{\gamma_{i}\}$.
	We just take the AM with QP as an example and set $V=2$ and $p=50$.
	In Fig. \ref{autocorr_LE} (a)-(c), 
	the autocorrelation for cross-section size $L=12$ have a tendency to go beyond the 95$\%$ confidence interval with increasing the transmission length $L_z$,
	which suggests the correlation would be strong for $\{\gamma_{i}\}$ with larger $L_z$. 
	%The standard error of autocorrelation reduces to $1/\sqrt{N}$ under the assumption the series is completely random. 
	%If the transmission length $L_z$ becomes smaller, the correlation could be within the 95$\%$ confidence interval.
	This means $L_z$ should be small if we want to guarantee that there is no correlation.
	However, to extract the critical exponent precisely from localization length, 
	the transmission length $L_z$ should be large enough, such as $10^7$.
	Then the correlation of the Lyapunov exponent will limit the precise estimations of 
	the critical exponents in quasiperiodic systems.
	We also note that a larger cross-section size $L=16$ could reduce the correlation [Fig. \ref{autocorr_LE}(d)], 
	but the cost of computation will increase as $L^7$.
	\bibliography{paper}
	
\end{document}